\documentclass[prb,twocolumn,superscriptaddress]{revtex4}

\usepackage{graphicx}
\usepackage{comment}
\usepackage{amsmath,amssymb,amsthm} 
\usepackage{verbatim}
\usepackage{float}
\usepackage{color}

\setcitestyle{super}

\usepackage[sort&compress]{natbib}

\begin{document}

\title{Effect of Energy Polydispersity on the Nature of \\ Lennard-Jones Liquids}
\author{Trond S. Ingebrigtsen}
\email{trond@iis.u-tokyo.ac.jp}
\affiliation{Institute of Industrial Science, University of Tokyo, 4-6-1 Komaba, Meguro-ku, Tokyo 153-8505, Japan}
\author{Hajime Tanaka}
\email{tanaka@iis.u-tokyo.ac.jp}
\affiliation{Institute of Industrial Science, University of Tokyo, 4-6-1 Komaba, Meguro-ku, Tokyo 153-8505, Japan}
\date{\today}

\begin{abstract}
In the companion paper [T. S. Ingebrigtsen and H. Tanaka, J. Phys. Chem. B 119, 11052 (2015)] the effect of size polydispersity on the nature of Lennard-Jones (LJ) liquids, which represent most molecular liquids 
without hydrogen bonds, was studied. More specifically, it was shown that even highly size polydisperse LJ liquids are Roskilde-simple (RS) liquids. RS liquids are liquids with strong correlation between constant volume equilibrium fluctuations of virial and potential energy and are simpler than other types of liquids. Moreover, it was shown that size polydisperse LJ liquids have isomorphs to a good approximation. Isomorphs are curves in the phase diagram of RS liquids along which structure, dynamics, and some thermodynamic quantities are invariant in dimensionless (reduced) units. In this paper, we study the effect of energy polydispersity on the nature of LJ liquids. We show that energy polydisperse LJ liquids are RS liquids. However, a tendency of particle segregation which increases with the degree of polydispersity leads to  a loss of strong virial-potential energy correlation, but is mitigated with increasing temperature and/or density. Isomorphs are a good approximation also for energy polydisperse LJ liquids, although particle-resolved quantities display a somewhat poorer scaling compared to the mean quantities along the isomorph. 
\end{abstract}

\maketitle

\section{Introduction}

Fluids showing a dispersion in a variable characterizing the constituent molecules are called polydisperse fluids\cite{bagchi2012,wolynes2012}. The dispersed variable could 
be the size of the molecules, but also their shape, charge, and more. A binary mixture comprised of two different molecular sizes is a simple example of 
a polydisperse fluid. However, in general the fluid can consist of numerous components. Bitumen is an example of a highly dispersed fluid 
with millions of components and constitutes the ``'glue'' part of asphalt owing to the unique and perplexing properties obtained from 
the highly polydisperse and viscous mixture\cite{hansen2013bit}.

Polydisperse fluids are interesting from both a theoretical and an experimental perspective\cite{dickinson1978,blum1979,salacuse1982,ginoza1997,sollich2002,fasolo2003,jacobs2013,
frenkel1986,kofke1986,wilding2005,wilding2005b,wilding2006,kawasaki2007,sarkar2013,sarkar2014,koningsveld1971,cowell1982,weeks2000,auer2001,watanabe2008,sacanna2013,palberg2014}. In 
experiments, for instance, polydispersity can be used as a tool to avoid crystallization and thus facilitate the study of highly viscous fluids\cite{weeks2000,leocmach2012}. The reason for the latter is the higher 
nucleation barriers of the supercooled polydisperse fluid\cite{auer2001} due to the increase of frustration against crystallization \cite{tanaka_review,mathieu_russo_tanaka}. Theoretically, polydisperse fluids pose a 
challenge as new perplexing phenomena, e.g., fractionation\cite{kofke1986,sollich2002,fasolo2003,wilding2005b,wilding2006} can occur.
Fractionation is the phenomenon that takes place when the fluid (or solid) forms additional phases, each showing a unique distribution. In fact, describing this 
phenomenon correctly has led to the creation of new theories as well as new simulation methods\cite{kofke1986,evans1999,fasolo2003,wilding2005b,wilding2006}.

In a recent paper\cite{ingebrigtsen2015} we studied the effect of size polydispersity on the property of being a so-called Roskilde-simple (RS) liquid\cite{paper0,paper1,paper2,paper3,paper4,paper5,prx}. 
RS liquids are characterized by having strong correlations between virial $W$ and potential energy $U$ equilibrium fluctuations at 
constant volume $V$. The virial-potential energy correlation is quantified via the correlation coefficient $R$ given by 

\begin{equation}\label{Rcor}
  R = \frac{\langle \Delta W \Delta U \rangle}{\sqrt{\langle (\Delta W)^{2} \rangle}\sqrt{ \langle (\Delta U)^{2} \rangle}},
\end{equation}
in which $\Delta$ denotes deviation from the mean values, $\langle ... \rangle$ provides \textit{NVT} ensemble averages, 
and $-1$ $\leq$ $R$ $\leq$ $1$. RS liquids are defined pragmatically\cite{paper1} by requiring $R$ $\geq$ $0.90$ and $R$ depends on the state point. 

The class of RS liquids is believed to include most or all van der Waals and metallic liquids, but to exclude most 
or all covalent-bonding, hydrogen-bonding, strongly ionic, and dipolar liquids\cite{paper1,prx}. RS liquids are simpler than other types of liquids displaying, e.g., 
isomorphs\cite{paper4,moleculesisomorphs}. Isomorphs are curves in the phase diagram of RS liquids along which structure, dynamics, and some 
thermodynamic quantities are invariant in dimensionless (reduced) units; reduced units refer in this case to macroscopic quantities, such as $T$ and $\rho^{-1/3}$, rather than the usual approach in simulations via microscopic quantities (see Ref. \onlinecite{paper4} for details).

In the companion paper\cite{ingebrigtsen2015}, we demonstrated that even highly size polydisperse Lennard-Jones (LJ) liquids are RS liquids. Furthermore, it was shown 
that these liquids have isomorphs to a good approximation. Experimental liquids, however, are often dispersed in additional variables besides their size. As an example: Colloidal fluids 
can have distribution in their size as well as their charge\cite{largo2006}; polymer fluids need both a size and energy distribution to be described correctly in simulations\cite{stapleton1988}. In this 
paper we study energy polydisperse LJ liquids as a first step in understanding how these additional factors affect polydisperse RS liquids. 
In experiments, a polydispersity in the refractive index of the particles should provide a similar situation.

An energy distribution has in previous studies been applied together with, e.g., a distribution in the molecular size\cite{kofke1986,stapleton1988}, and only
very recently been studied isolated for its interesting phase behavior \cite{shagolsem2015a,shagolsem2015b,rabin2016}. To fully unravel 
the relative effects of polydispersity on strong virial-potential energy correlation  we focus, in this study, on 
energy polydisperse LJ liquids \cite{shagolsem2015a}. Although this system is less experimentally realizable than size polydisperse systems, we believe that the conclusions 
drawn from this kind of study are necessary for understanding the effect of polydispersity on the nature of LJ liquids.

The paper is organized as follows. Simulation and model details are presented in Sec. \ref{sim}. Section \ref{ros} provides a brief introduction to RS liquids and 
their simple properties. Section \ref{res} presents molecular dynamics (MD) computer simulation results. Finally, Sec. \ref{con} summarizes the paper and provides a brief outlook.

\section{Simulation and model details}\label{sim}

MD computer simulations are utilized to investigate the effect of energy polydispersity on 
strong virial-potential energy correlation in the \textit{NVT} ensemble\cite{nvttoxvaerd}. The RUMD package\cite{rumd} is used to simulate 
highly-efficient GPU MD. A system size of $N$ = 131072 particles is used 
in the current study, and finite-size effects were checked by simulating $N$ = 62500 and $N$ = 5000 particles at selected state points. Finite-size effects were observed only 
for the $N$ = 5000 system at very high polydispersities.

The energy polydisperse LJ liquid has pair potential

\begin{equation}\label{lj}
  v_{ij}(r) = 4\epsilon_{ij}\Big[\Big(\frac{\sigma}{r}\Big)^{12} - \Big(\frac{\sigma}{r}\Big)^{6}\Big],
\end{equation}
and total potential energy function $U$ $\equiv$ $\sum_{i<j}v_{ij}(r)$. In Eq. (\ref{lj}): $\sigma_{}$ and $\epsilon_{ij}$ set, respectively, the length 
and energy scale of the pair interaction between particle $i$ and particle $j$ ($i$,$j$ = $1$, ..., $N$). The pair interaction energy $\epsilon_{ij}$ is assigned according to 
the rule\cite{tildesley,shagolsem2015a}

\begin{align}
  \epsilon_{i j} = \sqrt{\epsilon_{i} \, \epsilon_{j}}.
\end{align}
The distributed variables $\epsilon_{i}$ and $\epsilon_{j}$ are chosen from a uniform distribution (i.e., a flat distribution) on the interval from $\epsilon^{min}$ to $\epsilon^{max}$. 
The units of the simulation are defined by setting the mean values
over the $N$ particles to unity, i.e., $\sigma_{\bar{N} \bar{N}} = 1$, $\epsilon_{\bar{N} \bar{N}} = 1$, $m_{\bar{N}}$ = 1. A truncated-and-shifted pair potential cutoff at $r_{c} = 2.5\sigma$ is 
applied. We follow the standard approach in the literature and define the dimensionless polydispersity index $\delta$ as the ratio of the distributed variable's standard deviation and mean\cite{sarkar2013}.
Polydispersities are studied in the interval from $\delta$ = 0\% to $\delta$ = 51.92\%.

Furthermore, but in lesser detail, we study a modified version of the energy polydisperse LJ liquid in which the repulsive force is identical to that of a single-component 
LJ liquid, i.e., the repulsive force is not modified by the energy polydispersity. The pair potential is given by

\begin{align}
  v_{ij}^{att}(r) & = v_{ij}(r)/ \epsilon_{ij} + (1-\epsilon_{ij}),\, r < 2^{1/6}\sigma\label{newlj1},\\
  v_{ij}^{att}(r) & = v_{ij}(r),\, r \geq 2^{1/6}\sigma.\label{newlj2}
\end{align}
Figure \ref{pair} displays the range of pair potentials for these two systems with $\delta$ = 51.92\% (i.e., $\epsilon^{min}$ = 0.1 and $\epsilon^{max}$ = 1.9).
The motivation for studying the latter system is that Eq. (\ref{lj}), by modifying the repulsive force, also introduces a size polydispersity into the system at very high 
polydispersities (see the black curve in Fig. \ref{pair}(a)). 
\newline \newline
\begin{figure}[H]
  \centering
  \includegraphics[width=65mm]{lj}
\end{figure}
\begin{figure}[H]
  \centering
  \includegraphics[width=65mm]{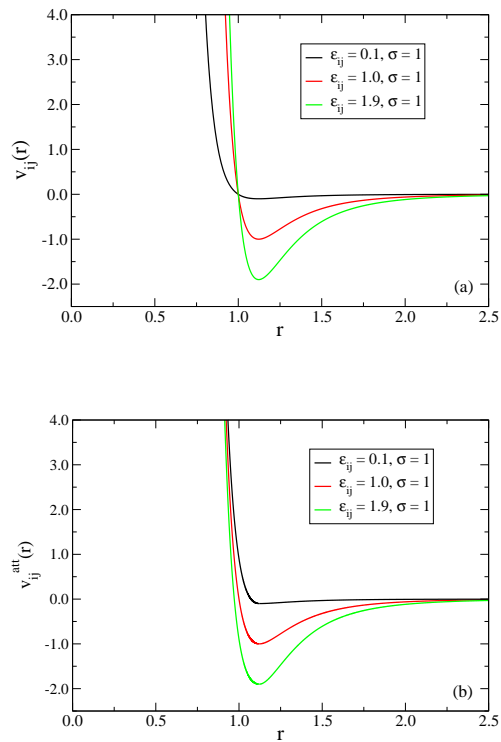}
  \caption{The range of pair potentials for energy polydisperse LJ liquids 
    with $\delta$ = 51.92\% (i.e., $\epsilon^{min}$ = 0.1 and $\epsilon^{max}$ = 1.9). (a) Pair potentials of Eq. (\ref{lj}). (b) Pair potentials of Eqs. (\ref{newlj1})-(\ref{newlj2}) in which the repulsive force is identical to that of a single-component LJ liquid.}
  \label{pair}
\end{figure}

\section{Roskilde-simple liquids}\label{ros}

This section provides a brief introduction to RS liquids and their associated simple properties\cite{paper0,paper1,paper2,paper3,paper4,paper5,prx}. A more detailed 
review is given in Ref. \onlinecite{dyre2014}. RS liquids were initially known as strongly-correlating liquids, but due to the confusion with strongly-correlated electron 
systems the name was changed in 2011 to RS liquids\cite{prx}. RS liquids are defined via the correlation coefficient

\begin{equation}\label{Rcor1}
  R = \frac{\langle \Delta W \Delta U \rangle}{\sqrt{\langle (\Delta W)^{2} \rangle}\sqrt{ \langle (\Delta U)^{2} \rangle}},
\end{equation}
to be whenever $R$ $\geq$ 0.90. $R$ varies throughout the phase diagram for a given liquid, but is usually observed to be high in most of the liquid part of the phase diagram for a RS liquid. 
In fact, strong correlation is also observed in the crystal phase showing higher correlation coefficients than in the liquid\cite{crystals}. Inverse power-law fluids 
interacting via $r^{-n}$ pair potentials are perfectly correlating with $R$ = 1 since $\Delta W = (n/3)\Delta U$.

Model systems identified to belong to RS liquids\cite{paper0,paper1,paper2,coslovich1,coslovich2,moleculeshidden}
include the standard single-component LJ liquid, the Kob-Andersen binary LJ mixture\cite{ka1}, 
the asymmetric dumbbell model\cite{moleculeshidden}, the Lewis-Wahnstr{\"o}m OTP model\cite{otp1}, and many more.
Strong correlation has also been experimentally verified for several van der Waals liquids\cite{gammagamma,roed2013,wence2014}.

RS liquids are characterized by having isomorphs in the phase diagram\cite{paper4,moleculesisomorphs}. Consider 
two state points in a liquid's phase diagram with density and temperature ($\rho_{1}$, $T_{1}$) and ($\rho_{2}$, $T_{2}$). These two state points are 
defined to be isomorphic if the following holds: Whenever micro-configurations of 
state point ($1$) and of state point ($2$) have the same reduced coordinates ($\textbf{R}$ is a 3$N$-dimensional configurational-space vector and $\rho \equiv N/V$),

\begin{equation}
  \rho_{1}^{1/3}\textbf{R}^{(1)} = \rho_{2}^{1/3} \textbf{R}^{(2)}, 
\end{equation}
these two configurations have proportional Boltzmann factors, i.e.,

\begin{equation} \label{defiso}
  e^{-U(\textbf{R}^{(1)})/k_{B}T_{1}} = C_{12} \, e^{-U(\textbf{R}^{(2)})/k_{B}T_{2}}.
\end{equation}
An isomorph is then defined as a continuous curve of state points for which all state points are pairwise isomorphic. $C_{12}$ is a constant and depends only on the 
state points ($1$) and ($2$), but not on their micro-configurations. Invariants along the isomorph include the structure and dynamics in dimensionless (reduced) units, but also several 
thermodynamics quantities such as the excess entropy $s_{ex}$ $\equiv$ $S_{ex}/N$ (with respect to an ideal gas).

RS liquids are simpler than other types of liquids\cite{prx,thermoscl}. Temperature, for example, factors\cite{thermoscl} into a product of a function of density and a function of excess entropy 

\begin{equation}
  T = h(\rho)f(s_{ex}). 
\end{equation}
The function $h(\rho)$ inherits the analytical structure of the potential energy function\cite{thermoscl} and can be used to generate 
isomorphs in simulations by keeping $h(\rho)/T$ constant since $s_{ex}$ is constant along an isomorph (see also Eq. (\ref{h}) in the next section). 
Previous investigations have also detailed that 
RS liquids and isomorphs are relevant for nanoconfined liquids\cite{ingebrigtsen2013,ingebrigtsen2014}, out-of-equilibrium liquids\cite{sllod}, 
polymer fluids\cite{veldhorst2014}, crystals\cite{crystals}, and more.

Recently, the concept of isomorphs was reformulated\cite{thomas2014}. The isomorphic condition now reads
\begin{equation}
  U(\textbf{R}_{a}) < U(\textbf{R}_{b}) \Rightarrow U(\lambda\textbf{R}_{a}) < U(\lambda\textbf{R}_{b}),
\end{equation}
where $\lambda$ is a scalar, and $\textbf{R}_{a}$ and $\textbf{R}_{b}$ are two different micro-configurations of a chosen state point. This definition 
encapsulates the previous definition of isomorphs (Eq. (\ref{defiso})) as a first-order 
approximation\cite{thomas2014}. Although the reformulation is a minor correction of the isomorph theory, it has the consequence that the 
constant-volume heat capacity is not (formally) invariant along an isomorph\cite{paper4,thomas2014}.

\section{Results}\label{res}

In this section we study energy polydisperse LJ liquids by varying the degree of polydispersity $\delta$. The effect of polydispersity 
on strong virial-potential correlation is studied first and subsequently isomorphs of 
energy polydisperse LJ liquids are investigated.

\subsection{Virial-potential energy correlation}\label{vircor}

The correlation coefficient $R$ and density-scaling exponent\cite{thermoscl} $\gamma$ $\equiv$ $\langle \Delta U \Delta W \rangle/\langle (\Delta U)^{2}\rangle$ 
are presented in Fig. \ref{R} as functions of polydispersity $\delta$ for the energy polydisperse LJ system (Eq. (\ref{lj})). The following state points are studied in the figure:

\begin{enumerate}
\item Three different temperatures $T$ = 0.70, 1.00, 4.00 with $\rho$ = 0.85 (black curves). These state 
  points correspond to a path from the triple point into the fluid region at $\delta$ = 0\%. 
\item A state point with $\rho$ = 0.95 and $T$ = 1.50 (red curve). This state point is located close to the freezing line at $\delta$ = 0\%.
\item A state point with $\rho$ = 1.10 and $T$ = 0.70 (blue curve). A perfect FCC lattice was prepared, a random (energy) distribution in space introduced, and the crystal simulated.
\end{enumerate}
Each of the black curves displays a weakening of the strong virial-potential energy correlation with increasing polydispersity (Fig. \ref{R}(a)). In fact, the curve at $T$ = 0.70 shows a very 
strong effect of 
polydispersity on $R$; the liquid is no longer RS for polydispersities above $\delta$ $\approx$ 30\%. For $T$ = 1.00, the liquid becomes non-RS at $\delta$ $\approx$ 40\%,
whereas all investigated state points remain RS at $T$ = 4.00. The weakening of the strong 
virial-potential energy correlation is thus mitigated with increasing temperature. A similar conclusion is reached when simultaneously increasing density and temperature (red curve).
The crystal state points display only a weak effect of polydispersity on $R$ (blue curve).
\newline \newline
\begin{figure}[H]
  \centering
  \includegraphics[width=70mm]{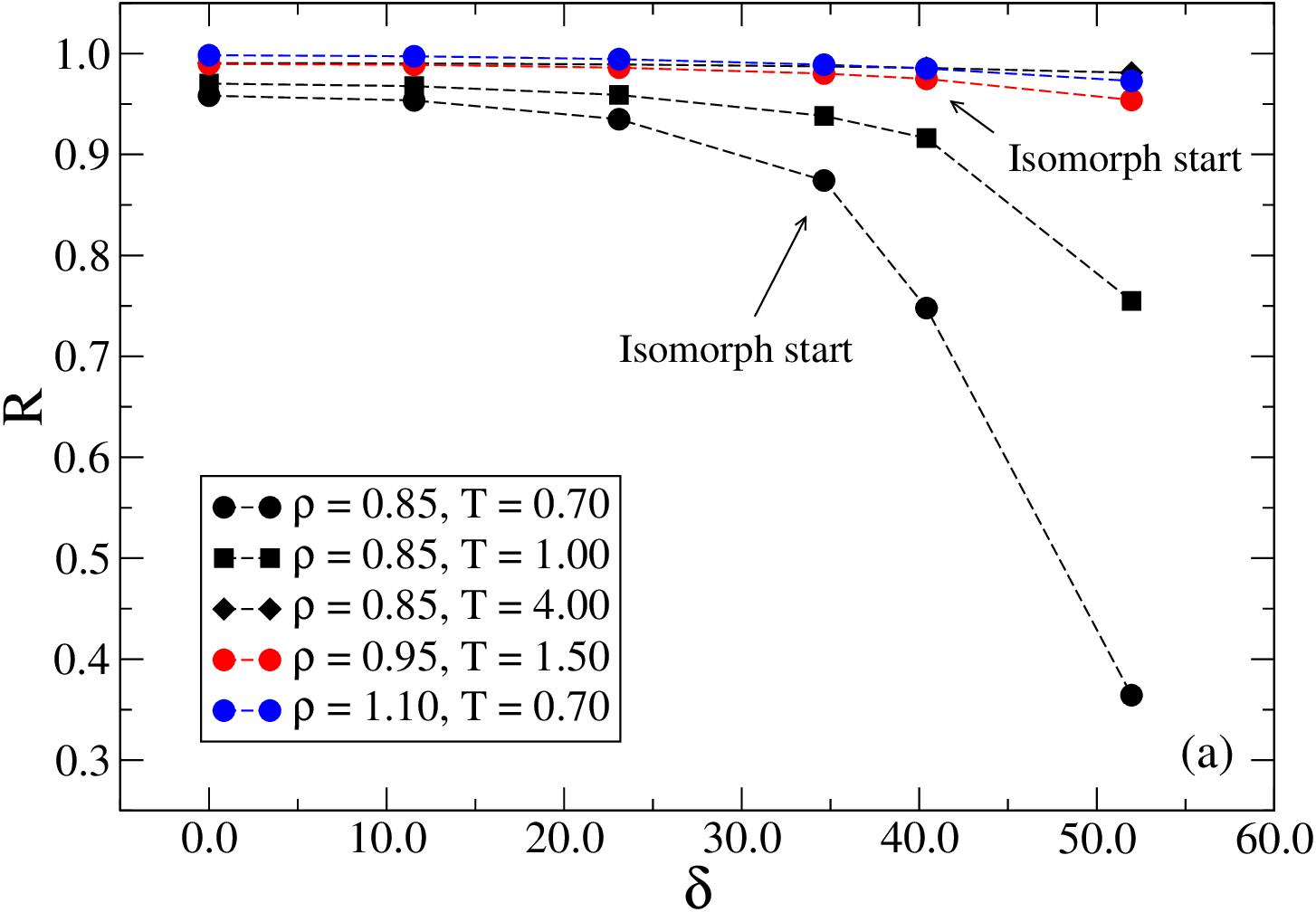}
\end{figure}
\begin{figure}[H]
  \centering
  \includegraphics[width=70mm]{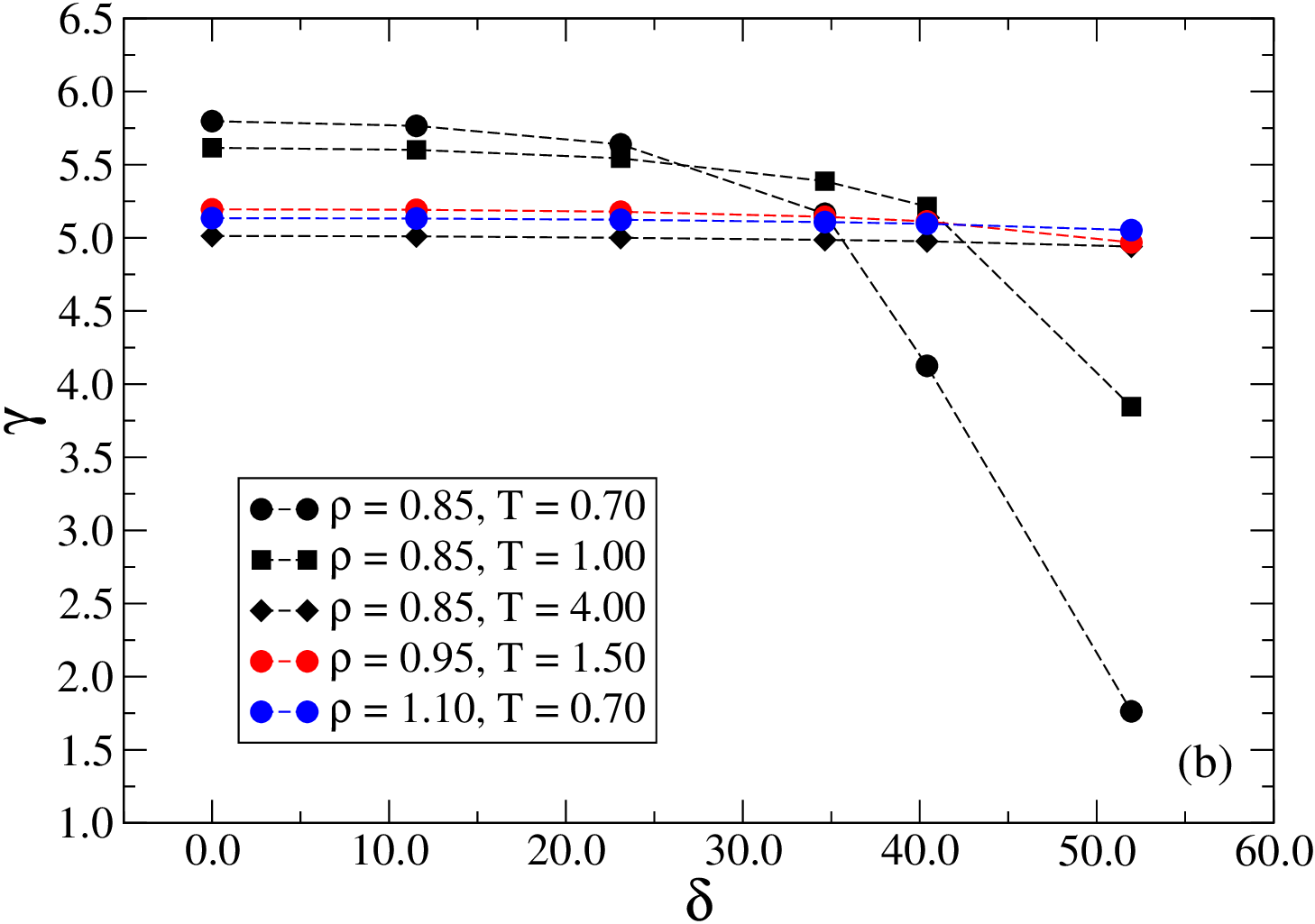}
  \caption{Effect of polydispersity $\delta$ on the correlation coefficient $R$ (Eq. (\ref{Rcor1})) and density-scaling exponent $\gamma$ $\equiv$ $\langle \Delta U \Delta W \rangle/\langle (\Delta U)^{2}\rangle$ for the energy polydisperse LJ system (Eq. (\ref{lj})). The paths studied are described in the text. The text ``Isomorph start'' refers to Sec. \ref{isosec}. (a) $R$ as function of $\delta$. (b) $\gamma$ as function of $\delta$. }
  \label{R}
\end{figure}
To obtain a better understanding of why $R$ decreases so dramatically as function of polydispersity, we show in Fig. \ref{sep} a snapshot of a configuration 
at $\rho$ = 0.85, $T$ = 0.70, and $\delta$ = 51.96\% with particles colored according to their $\epsilon_{i}$-values (a system with $N$ = 5000 is used for illustration). Figure \ref{sep} shows  
that the particles are segregated to a great extent leading to the loss of the strong virial-potential energy correlation. This observation is consistent with 
previous investigations of the phase behavior of energy polydisperse LJ systems\cite{shagolsem2015a,shagolsem2015b,rabin2016}.

The particle segregation is driven by the inclusion of high $\epsilon_{i}$-value particles, and can naturally 
be mitigated by increasing temperature and/or density and thus also explains the observed behavior
for $R$ as function of temperature and/or density (black and red curves). The particles of high $\epsilon_{i}$-values crystallize after simulating the system many millions of time steps. 

\begin{figure}[H]
  \centering
  \includegraphics[width=65mm]{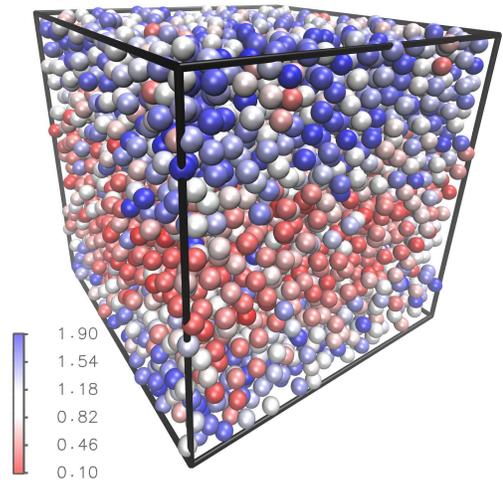}
  \caption{Snapshot of a configuration at $\rho$ = 0.85, $T$ = 0.70, and $\delta$ = 51.96\% (A system with $N$ = 5000 is used for illustration). Particles 
are colored according to their $\epsilon_{i}$-values ($\epsilon^{min}$ = 0.1 and $\epsilon^{max}$ = 1.9). Phase separation is observed. The particles of high $\epsilon_{i}$-values crystallize 
after simulating the system many millions of time steps.} 
\label{sep}
\end{figure}

\begin{figure}[H]
  \centering
  \includegraphics[width=70mm]{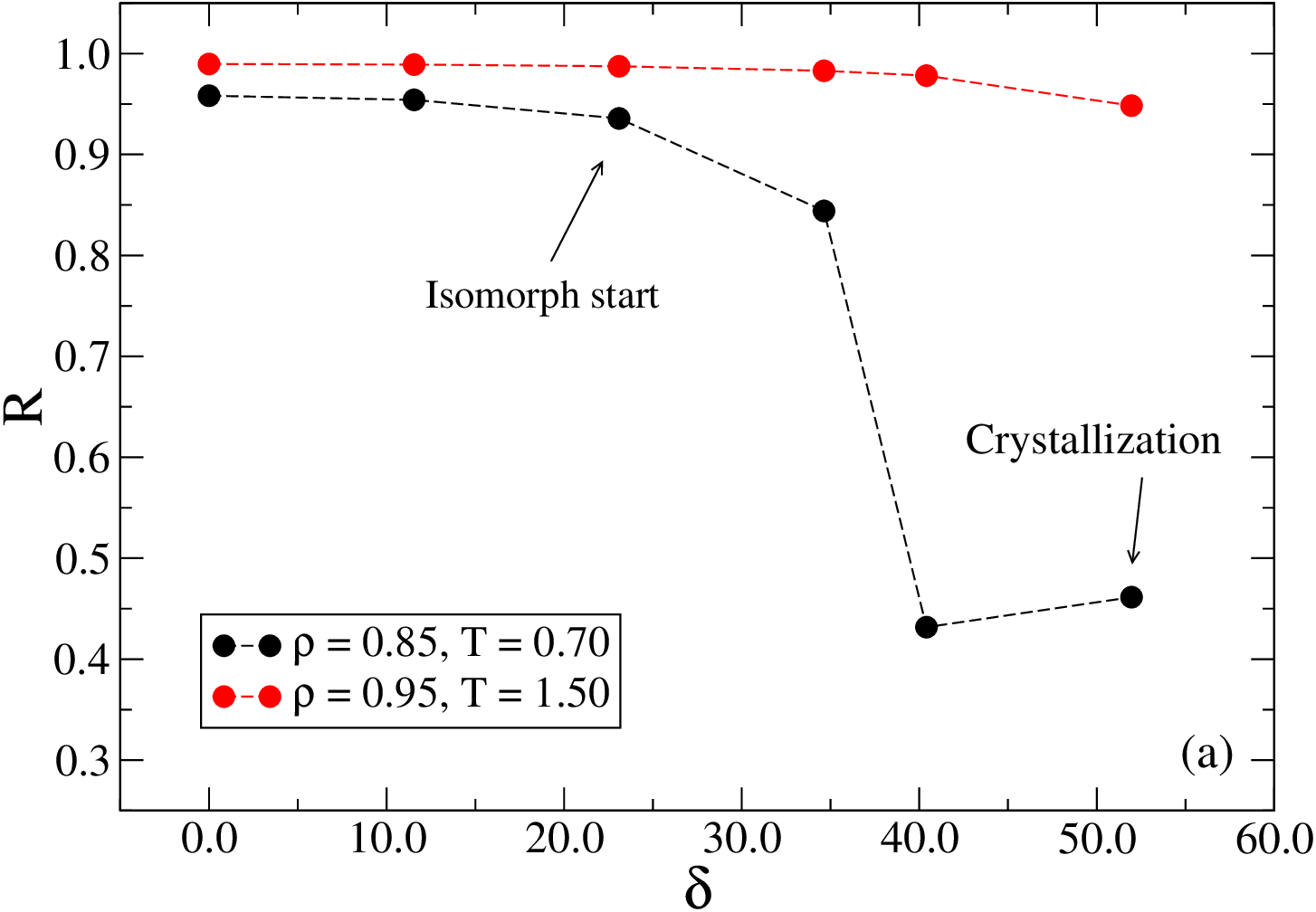}
\end{figure}
\begin{figure}[H]
  \centering
  \includegraphics[width=70mm]{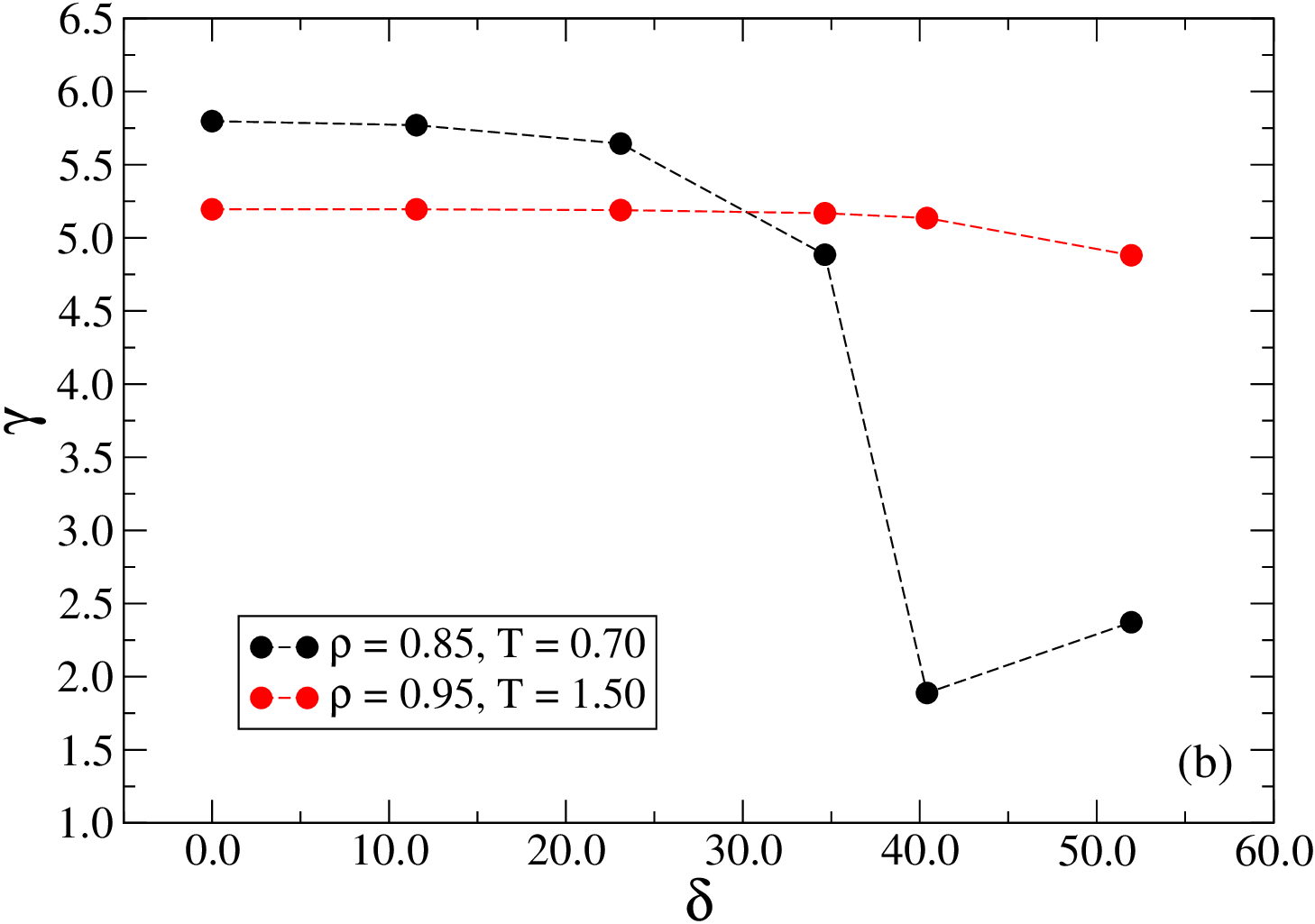}
  \caption{Effect of polydispersity $\delta$ on the correlation coefficient $R$ (Eq. (\ref{Rcor1})) and density-scaling exponent $\gamma$ $\equiv$ $\langle \Delta U \Delta W \rangle/\langle (\Delta U)^{2}\rangle$ for the modified energy polydisperse LJ system (Eqs.  (\ref{newlj1}) and (\ref{newlj2})). The text ``Isomorph start'' refers 
    to Sec. \ref{isosec}. (a) $R$ as function of $\delta$. (b) $\gamma$ as function of $\delta$. }
  \label{Rnew}
\end{figure}
Next, Fig. \ref{Rnew} shows the correlation coefficient $R$ and density-scaling exponent $\gamma$ along two paths (see previous description) for the modified 
energy polydisperse LJ system (Eqs.  (\ref{newlj1}) and (\ref{newlj2})). Up until a polydispersity of $\delta$ $\approx$ 30\% (black curve) $R$ shows almost identical behavior with the original 
system (Fig. \ref{R}(a)). However, after passing this value the correlation coefficient drops quickly for the modified system and crystallization occurs at $\delta$ = 52\% in the simulation time scale. The density-scaling exponent $\gamma$ 
shows a similar behavior as for the correlation coefficient. For the red curves only minor differences are noted between the two systems. The main differences between the original 
and modified LJ system thus first seem to appear when the original system becomes non-RS. 

We believe that the difference in phase behavior between the two systems originates from the polydispersity of the repulsive core size: 
For the potential given by Eq. (2), energy polydispersity automatically leads to a size polydispersity of the repulsive core (see Fig. 1(a)) which causes geometrical and structural frustration against crystallization\cite{mathieu_russo_tanaka}. On the other hand, for the potential given by Eqs. (4) and (5) energy polydispersity does not induce such a size polydispersity. 

\subsection{Isomorphs}\label{isosec}

Next we focus on isomorphs of energy polydisperse LJ liquids. The generation of isomorphs follows a 
procedure similar to that outlined in Ref. \onlinecite{ingebrigtsen2015} by keeping $h(\rho)/T$ constant (see also Sec. \ref{ros}). The function $h(\rho)$ is given by\cite{thermoscl,beyond}

\begin{align}\label{h}
  h(\tilde{\rho}) = (\gamma_{*}/2 - 1)\tilde{\rho}^{4} + (2 - \gamma_{*}/2)\tilde{\rho}^{2},
\end{align}
where $\tilde{\rho}$ $\equiv$ $\rho / \rho_{*}$,
$\rho_{*}$ is a chosen reference density, and $\gamma_{*}$ is the value 
of $\gamma$ = $\langle \Delta U \Delta W \rangle/\langle (\Delta U)^{2}\rangle$ obtained from the equilibrium fluctuations at $\rho_{*}$ (and $T_{*}$). 
The procedure is as follows:

\begin{enumerate}
\item A starting state point is chosen.
\item Density is varied by some percentage.
\item The temperature of the isomorphic state point is calculated from Eq. (\ref{h}) by keeping $h(\tilde{\rho})/T$ constant.
\item A simulation is performed at the predicted state point. The
  procedure is then repeated by varying the density by another percentage, i.e., step 2.
\end{enumerate}
Here and in the following we choose the reference density $\rho_{*}$ of Eq. (\ref{h}) to be that of the starting state point for the investigated isomorph\cite{thermoscl}.
We start the investigation by considering an isomorph for an energy polydisperse LJ liquid (Eq. (\ref{lj})) with polydispersity $\delta$ = 34.64\%. The starting state 
point is $\rho$ = 0.85 and $T$ = 0.70, and has $R$ $\approx$ 0.87 (see text in Fig. \ref{R}(a)). Although 
the starting state point is then a border-line RS liquid, this study can provide valuable insight into 
the extent to which particle segregation is isomorph invariant (see also Sec. \ref{vircor}). 

Radial distribution functions (RDFs) along the generated isomorph and an isotherm are displayed in Fig. \ref{rdf3d2}. The isotherm 
serves the purpose of a reference system, as the isomorph theory is approximate\cite{paper1,paper4}.
The average structure is to a good approximation invariant along the isomorph, but not on the isotherm where the liquid actually crystallizes.

\begin{figure}[H]
  \centering
  \includegraphics[width=70mm]{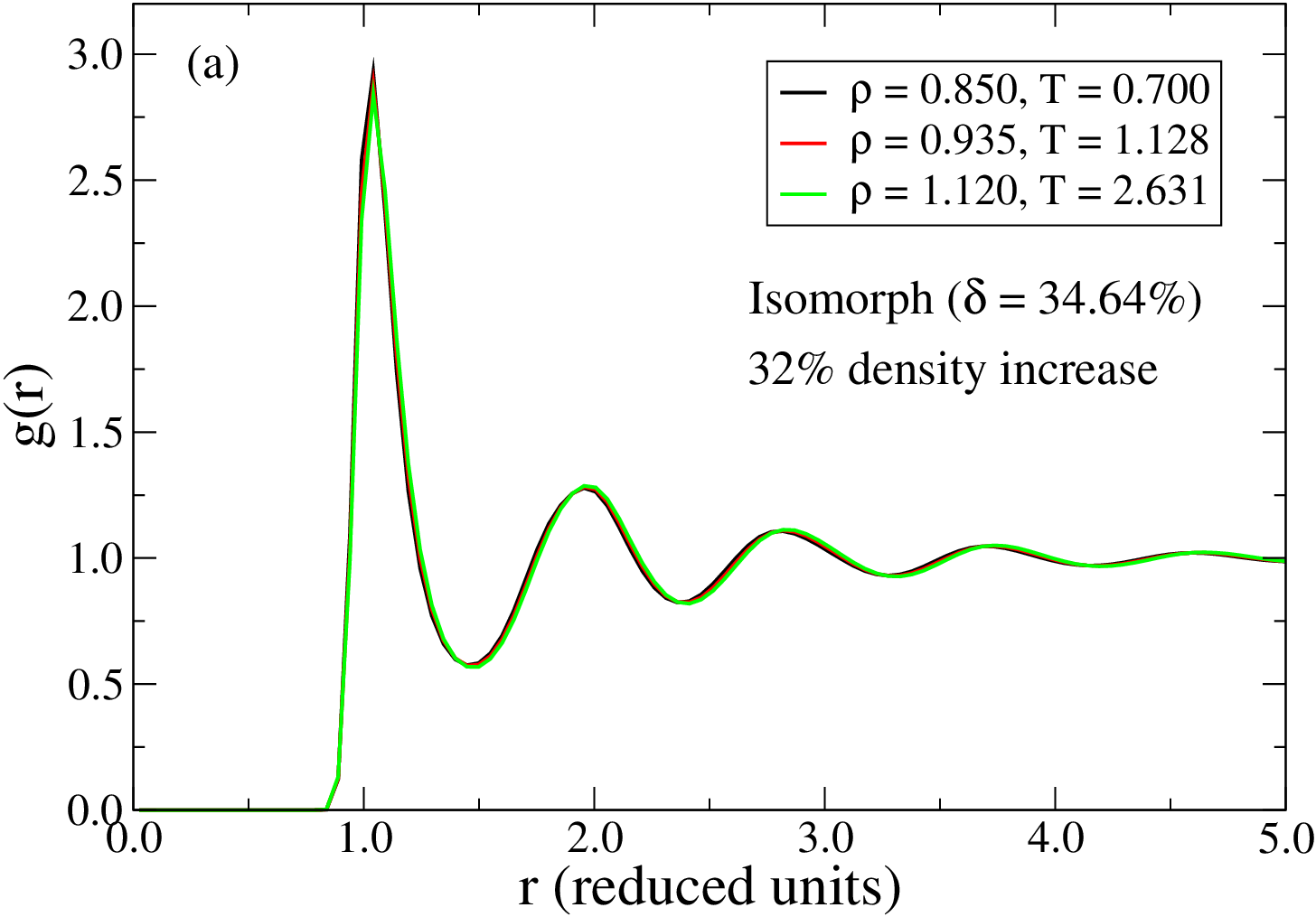}
\end{figure}
\begin{figure}[H]
  \centering
  \includegraphics[width=70mm]{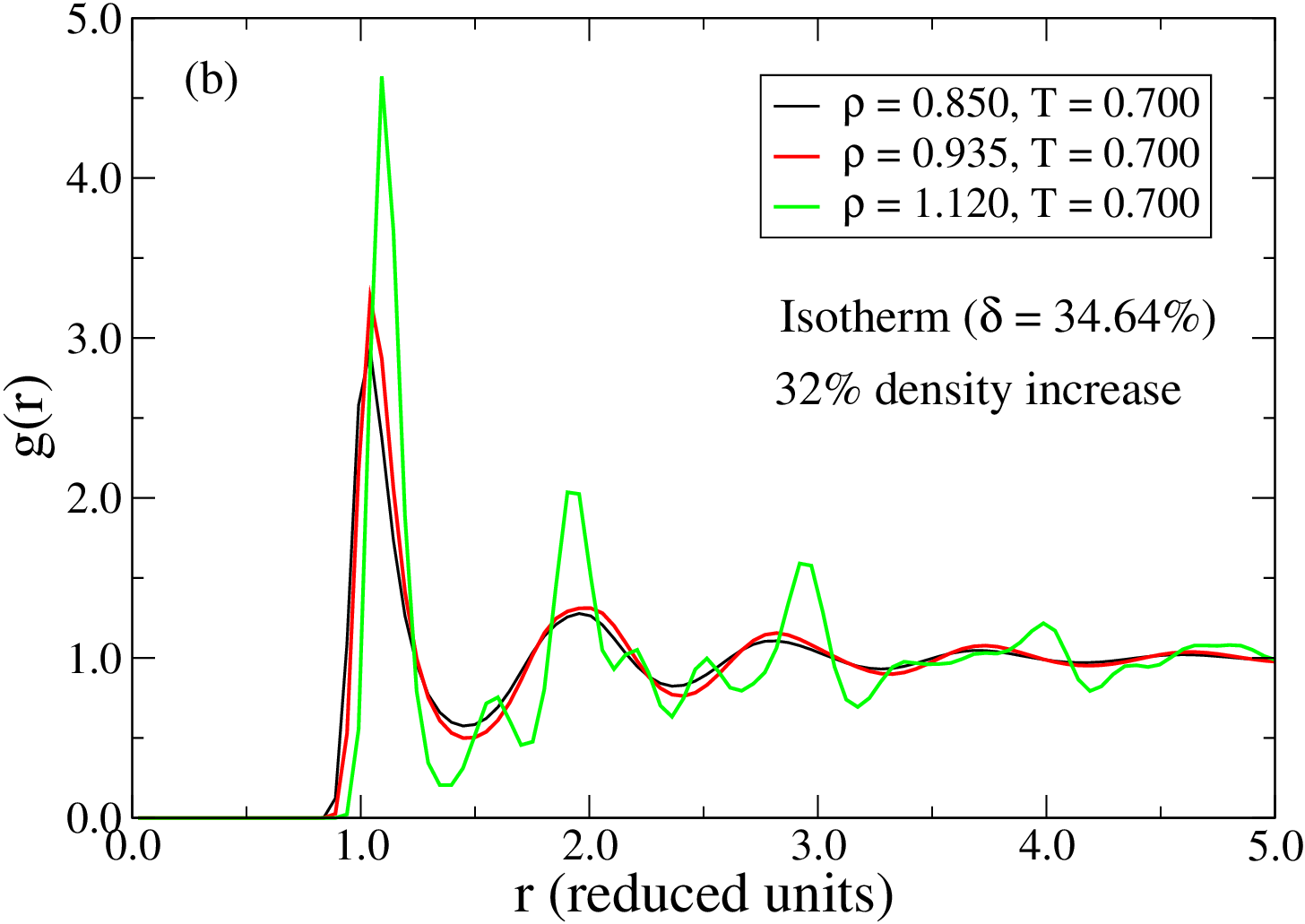}
  \caption{Radial distribution functions (RDFs) for an energy polydisperse LJ liquid at $\delta$ = 34.64\% (Eq. (\ref{lj})). (a) Along an isomorph. (b) Along an isotherm.}
  \label{rdf3d2}
\end{figure}
The dynamics for the same isomorph and isotherm is presented in Fig. \ref{msd3d2}. The mean-square displacements (MSDs) scale excellently along the 
isomorph, whereas the isotherm shows more than six orders-of-magnitude variation (Fig. \ref{msd3d2}(b)).

\begin{figure}[H]
  \centering
  \includegraphics[width=70mm]{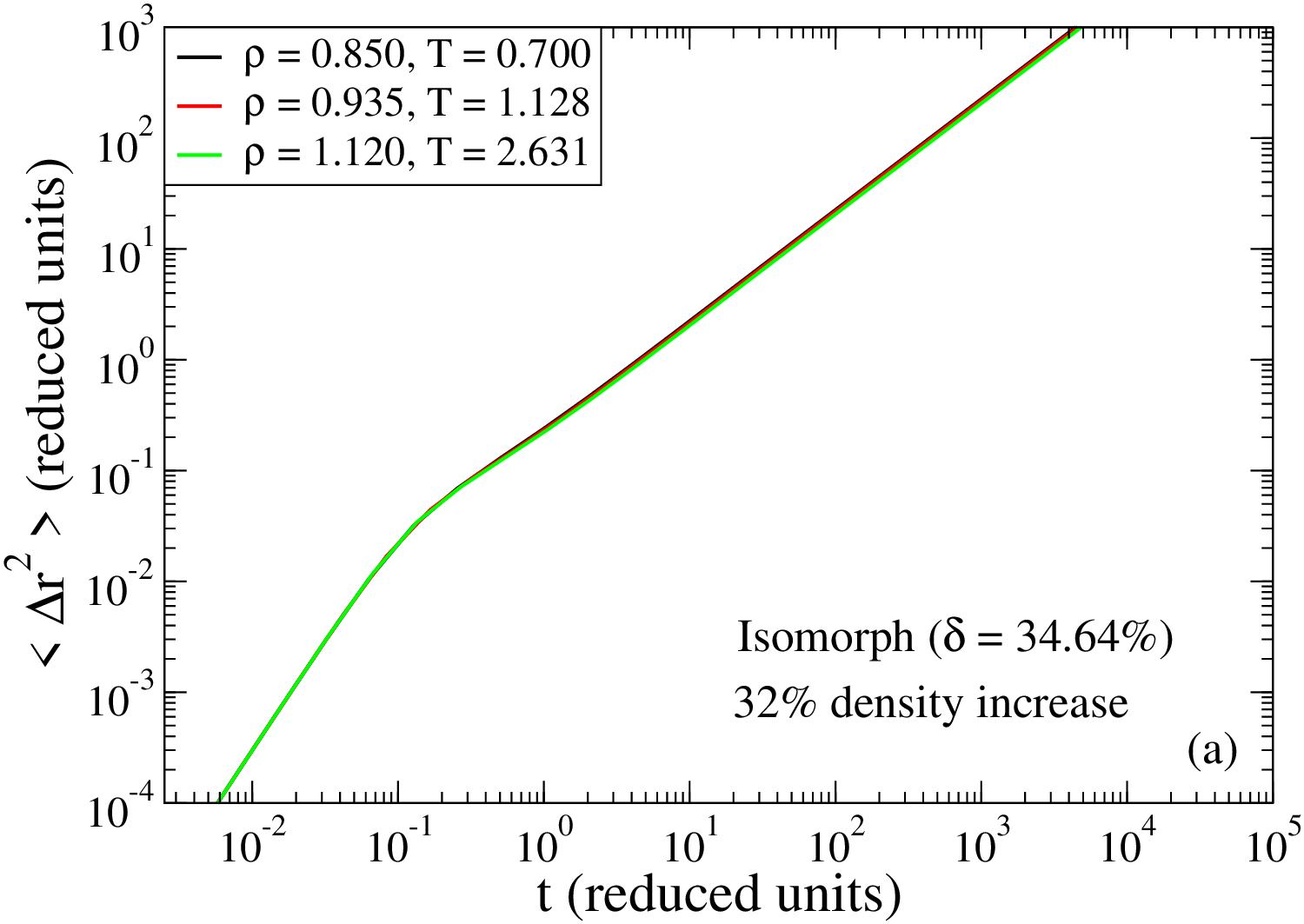}
\end{figure}
\begin{figure}[H]
  \centering
  \includegraphics[width=70mm]{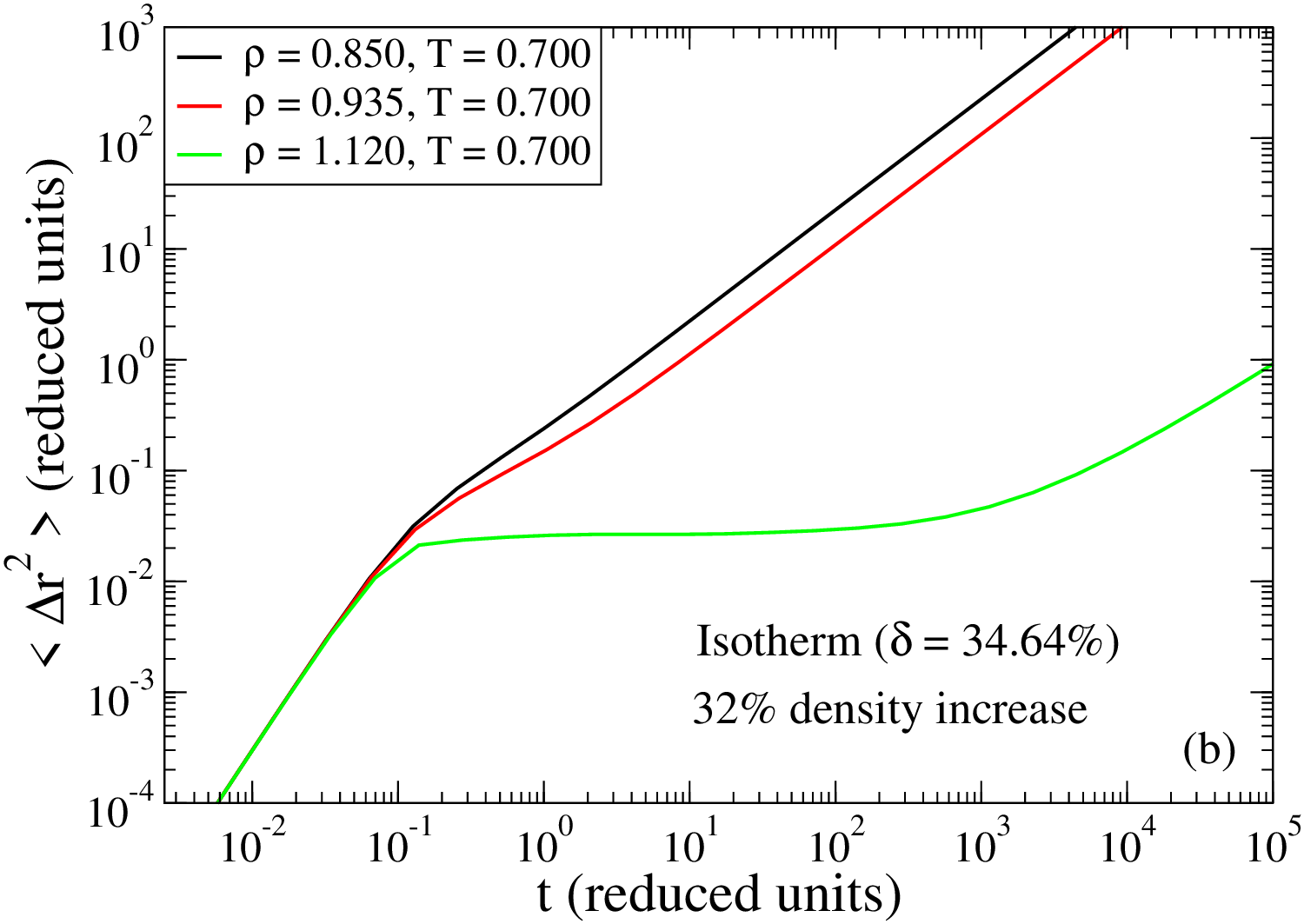}
  \caption{Mean-square displacements (MSDs) for the energy polydisperse LJ liquid of Fig. \ref{rdf3d2}. (a) Along the isomorph. (b) Along the isotherm. At the highest density (green curve)
  the isotherm crystallizes and gives rise to the very slow dynamics.}
  \label{msd3d2}
\end{figure}
Particle segregation is studied in Fig. \ref{sqf3d2} by probing the static structure factor (SSF) along the same isomorph and isotherm. The low $q$-values 
of the SSF display a poor invariance along the isomorph and indicate that the 
particle segregration is not fully isomorph invariant. The latter is expected as the particle segregration is the reason why $R$ decreases
with increasing polydispersity (see Sec. \ref{vircor}). The invariance along the isomorph is, however, much better than along the isotherm (Fig. \ref{sqf3d2}(b)).
\newline \newline
\begin{figure}[H]
  \centering
  \includegraphics[width=70mm]{isomorph2_sq}
\end{figure}
\begin{figure}[H]
  \centering
  \includegraphics[width=70mm]{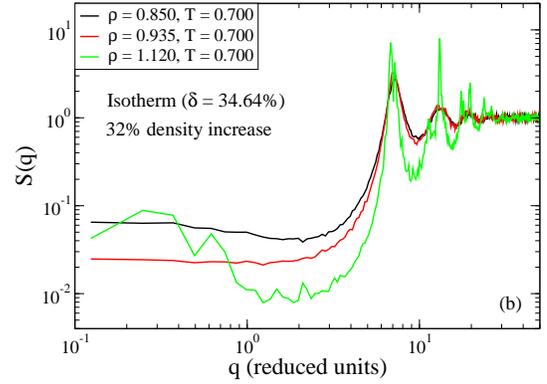}
  \caption{Static structure factors (SSFs) for the energy polydisperse LJ liquid of Figs. \ref{rdf3d2}-\ref{msd3d2}. (a) Along the isomorph. (b) Along the isotherm.}
  \label{sqf3d2}
\end{figure}
The fact that the particle segeration is not isomorph invariant can also be seen by probing particle-resolved quantities. Particle-resolved RDFs are shown in Fig. \ref{particle} where 
the dashed curves for clarity have been shifted with 0.5 in the $x$-direction. The 
RDFs of the low $\epsilon_{i}$-value particles scale well along the isomorph (full curves), but the RDFs of 
the high $\epsilon_{i}$-value particles scale poorly (dashed curves). In both cases, however, much better than along the isotherm (Fig. \ref{particle}(b)). 

\begin{figure}[H]
  \centering
  \includegraphics[width=70mm]{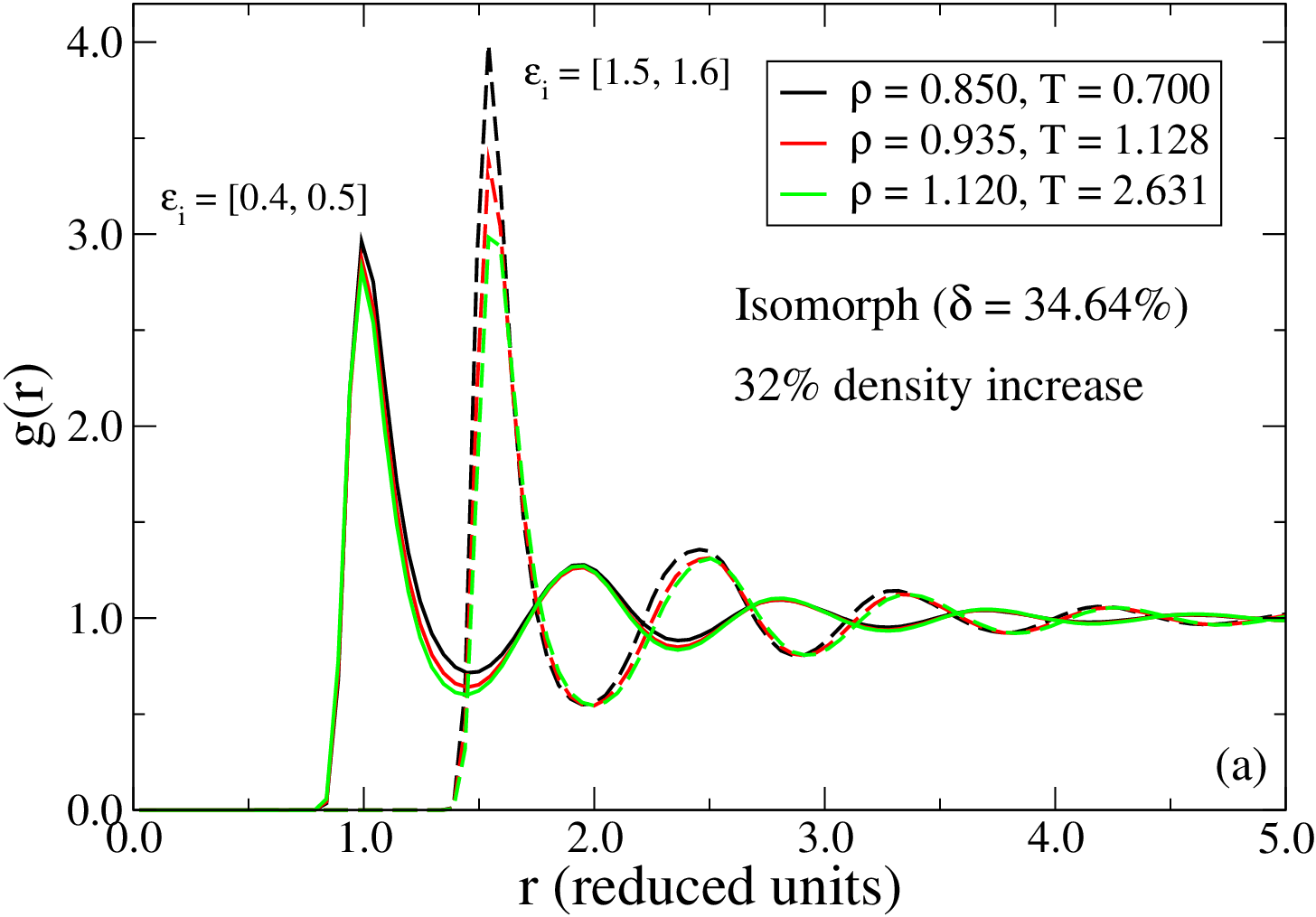}
\end{figure}
\begin{figure}[H]
  \centering
  \includegraphics[width=70mm]{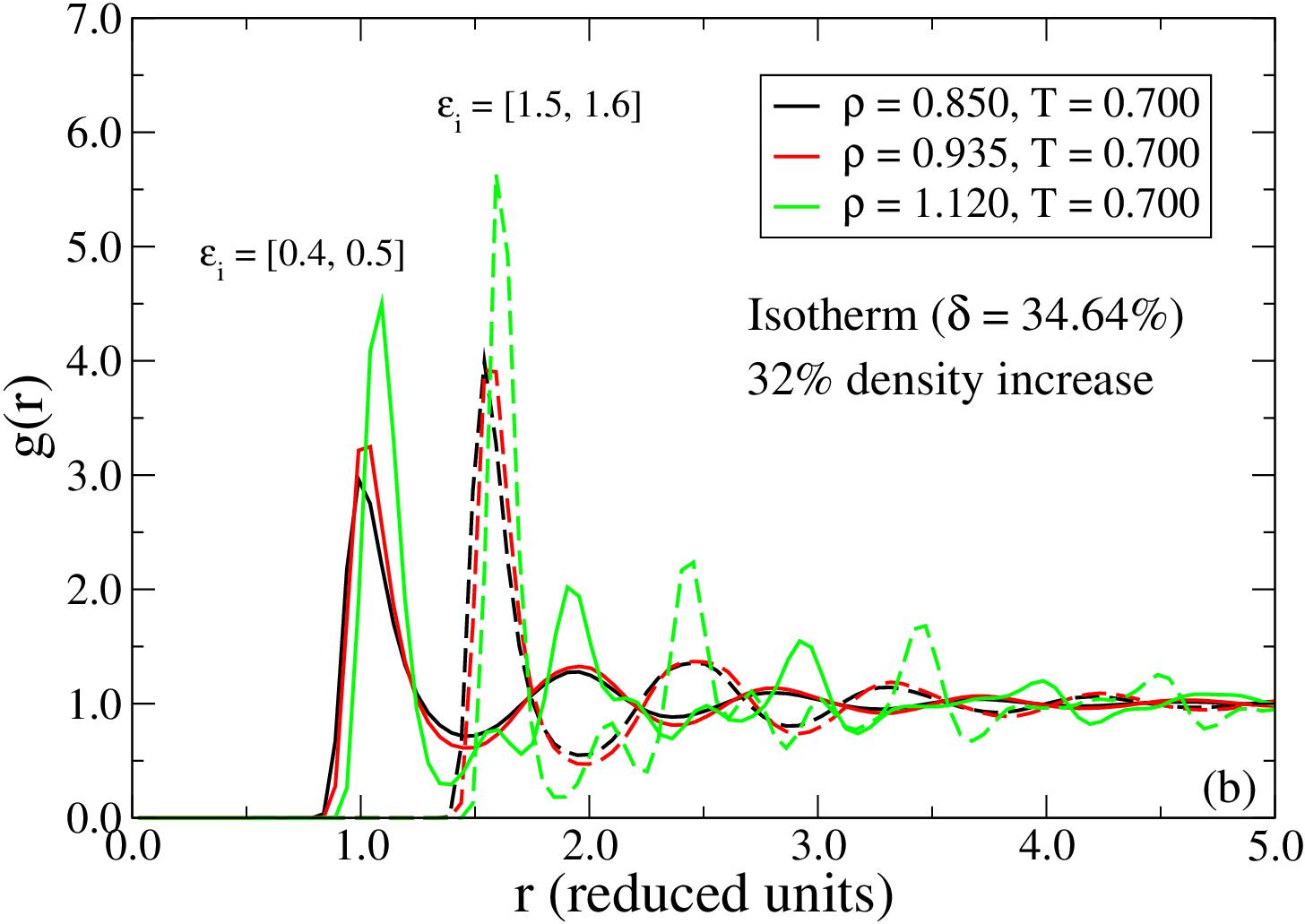}
  \caption{Particle-resolved RDFs for the energy polydisperse LJ liquid of Figs. \ref{rdf3d2}-\ref{sqf3d2}. The full curves use particle energies in the range $\epsilon_{i}$ = [0.4, 0.5], and 
  the dashed curves use $\epsilon_{i}$ = [1.5, 1.6]. The dashed curves have for clarity been shifted with 0.5 in the $x$-direction. (a) Along the isomorph. (b) Along the isotherm.}
\label{particle}
\end{figure}
The study is now focused on an isomorph for which the energy polydispersity is very high, i.e., $\delta$ = 40.41\% (Eq. (\ref{lj})). The motivation for studying such a high 
polydispersity is to carefully test the range of validity of isomorphs. The starting state point is $\rho$ = 0.95 and $T$ = 1.50 and has $R$ $\approx$ 0.97. 
Figure \ref{rdf3d1} displays RDFs along the isomorph and an isotherm. The RDFs scale excellently along the isomorph whereas the isotherm shows deviations; at the highest density due to 
crystallization during the simulation.

\begin{figure}[H]
  \centering
  \includegraphics[width=70mm]{isomorph1_rdf}
\end{figure}
\begin{figure}[H]
  \centering
  \includegraphics[width=70mm]{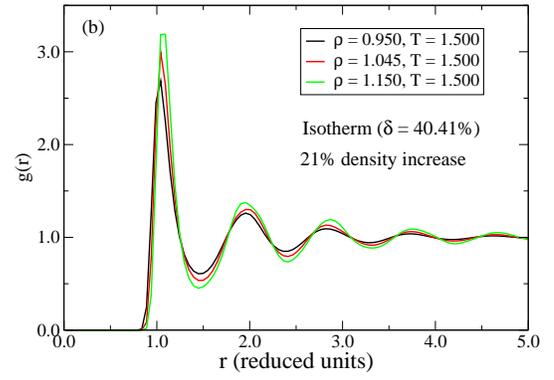}
  \caption{RDFs for an energy polydisperse LJ liquid at $\delta$ = 40.41\% (Eq. (\ref{lj})). (a) Along an isomorph. (b) Along an isotherm.}
  \label{rdf3d1}
\end{figure}
MSDs are displayed in Fig. \ref{msd3d1} and show perfect scaling along the isomorph but not along the isotherm, as expected. 
\newline \newline
\begin{figure}[H]
  \centering
  \includegraphics[width=70mm]{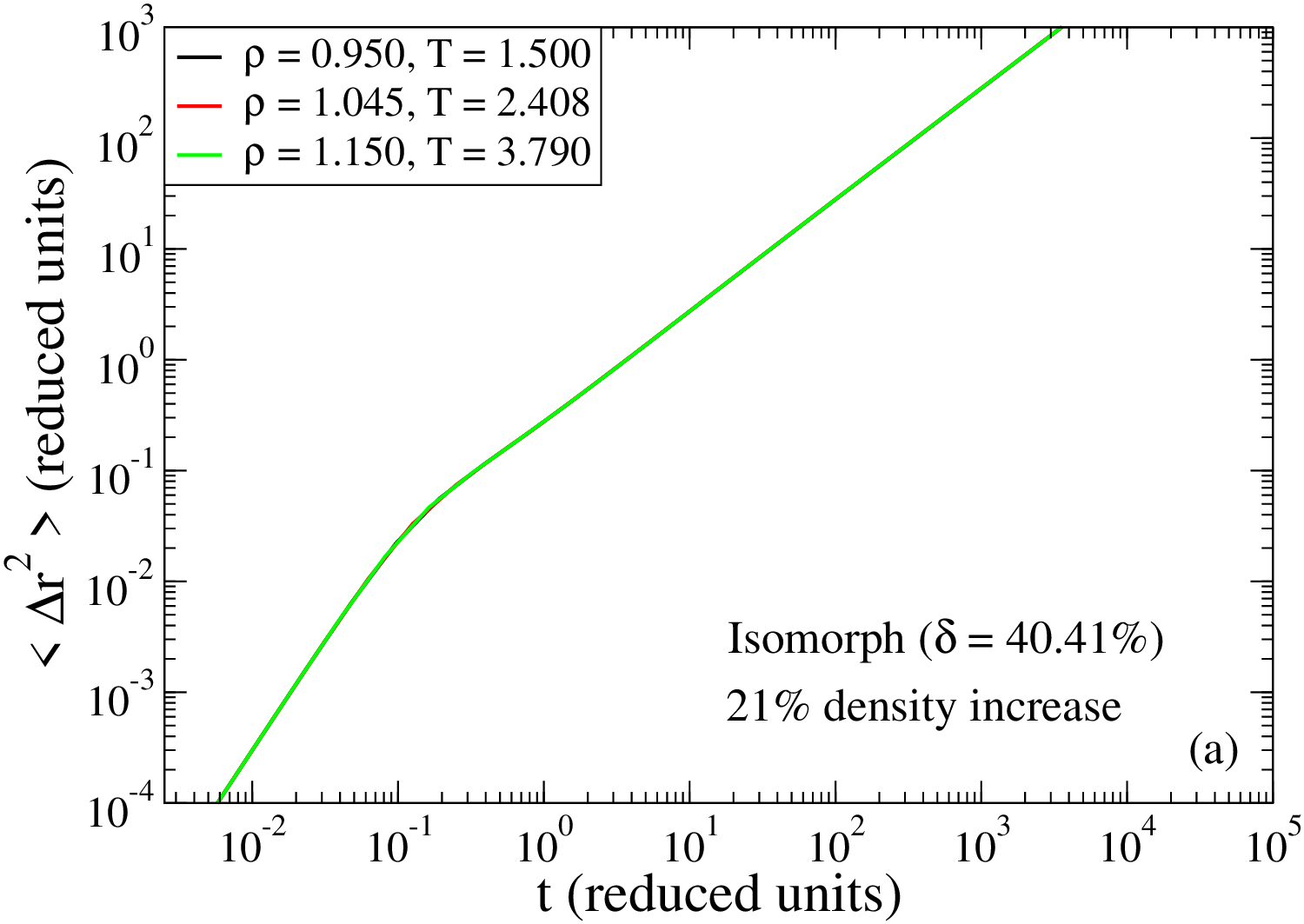}
\end{figure}
\begin{figure}[H]
  \centering
  \includegraphics[width=70mm]{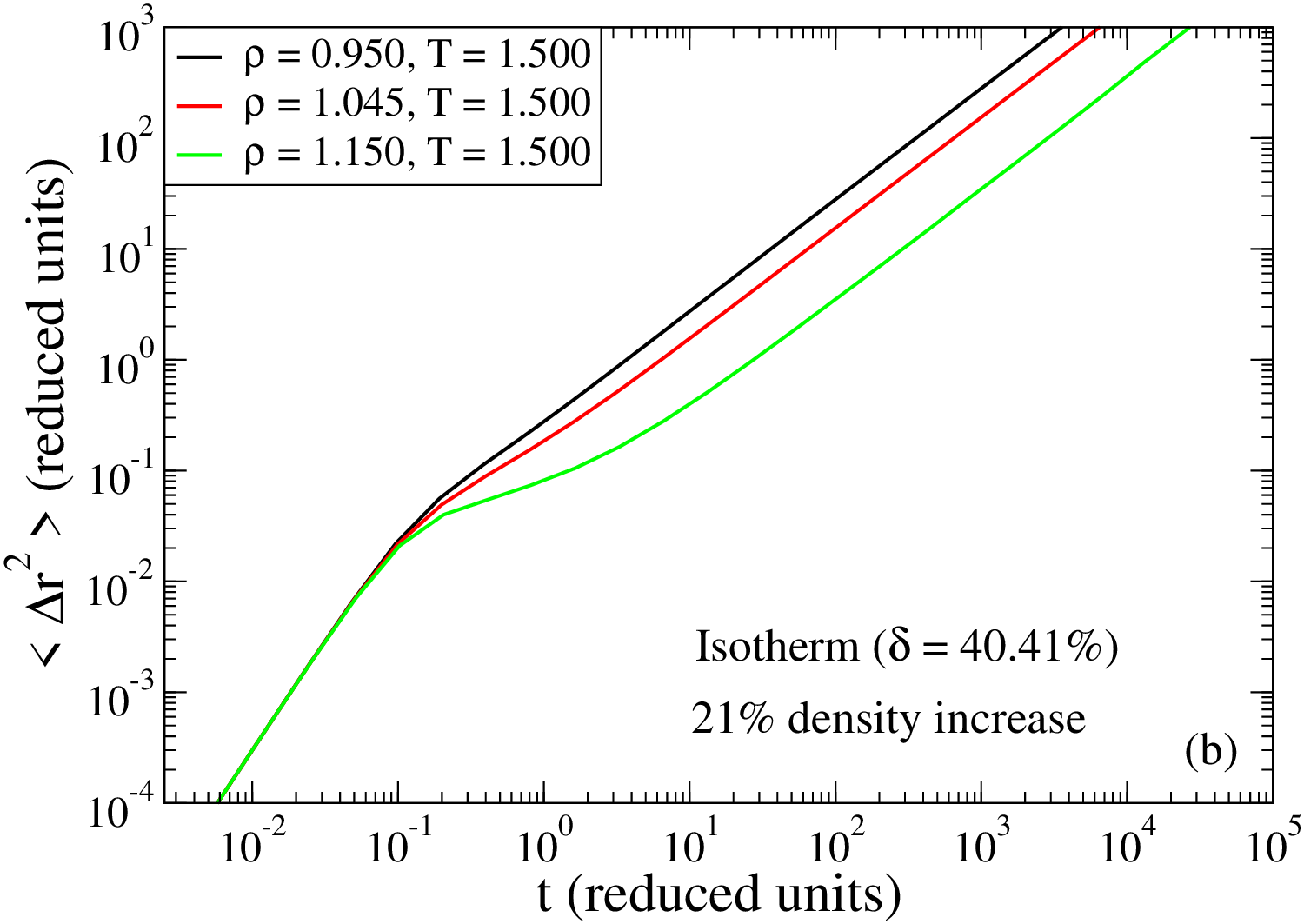}
  \caption{MSDs for the energy polydisperse LJ liquid of Fig. \ref{rdf3d1}. (a) Along the isomorph. (b) Along the isotherm.}
  \label{msd3d1}
\end{figure}
Figure \ref{sq3d1} presents SSFs along the same isomorph and isotherm. The isomorph displays again deviations in the low-$q$ part of the SSF. In this case, however, 
much smaller deviations than the previous isomorph as the starting state point of the isomorph has only weak particle segregation. The isotherm, on the other hand, shows very clear deviations, in particular
due to crystallization.
\newline \newline
\begin{figure}[H]
  \centering
  \includegraphics[width=70mm]{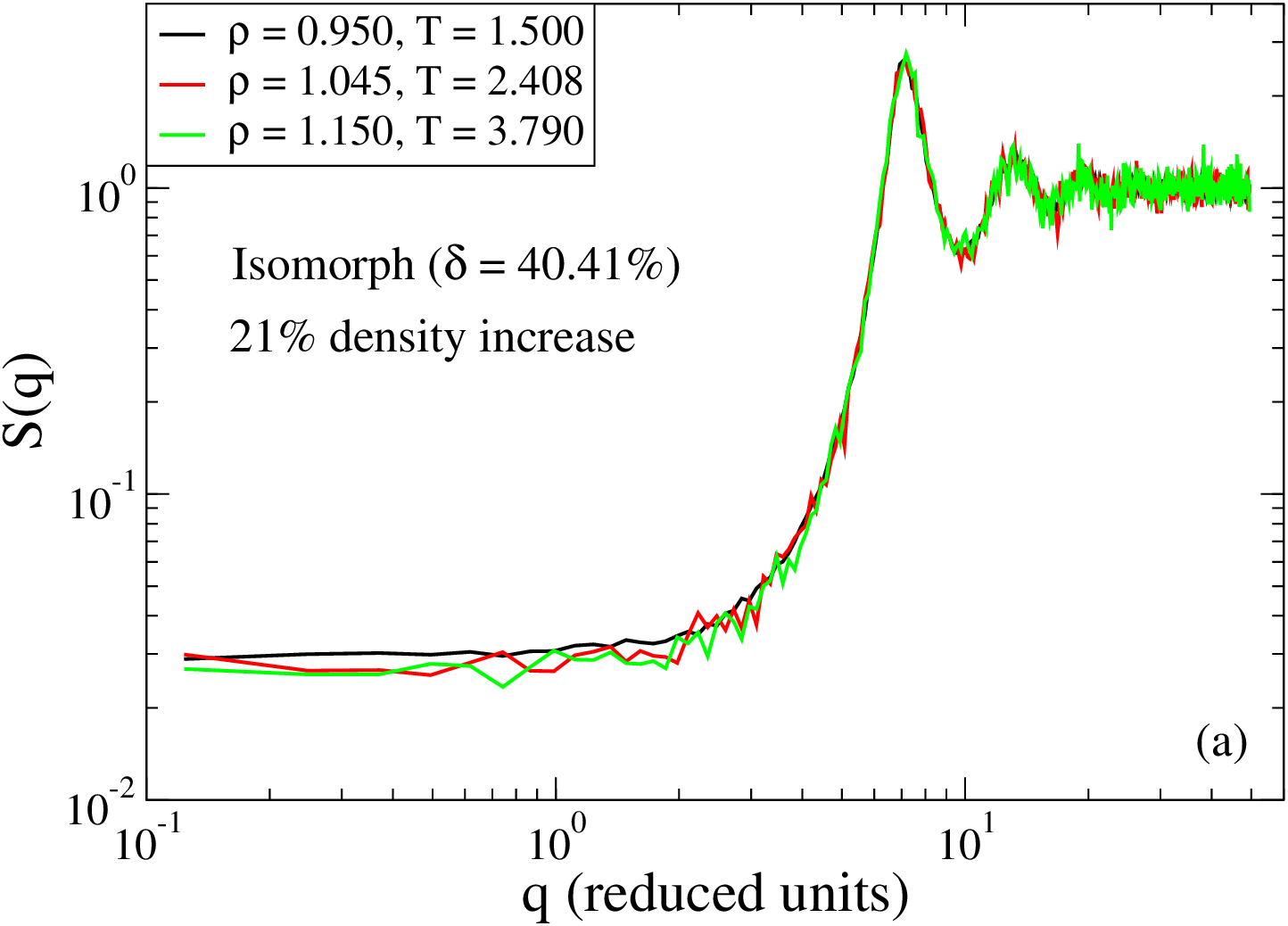}
\end{figure}
\begin{figure}[H]
  \centering
  \includegraphics[width=70mm]{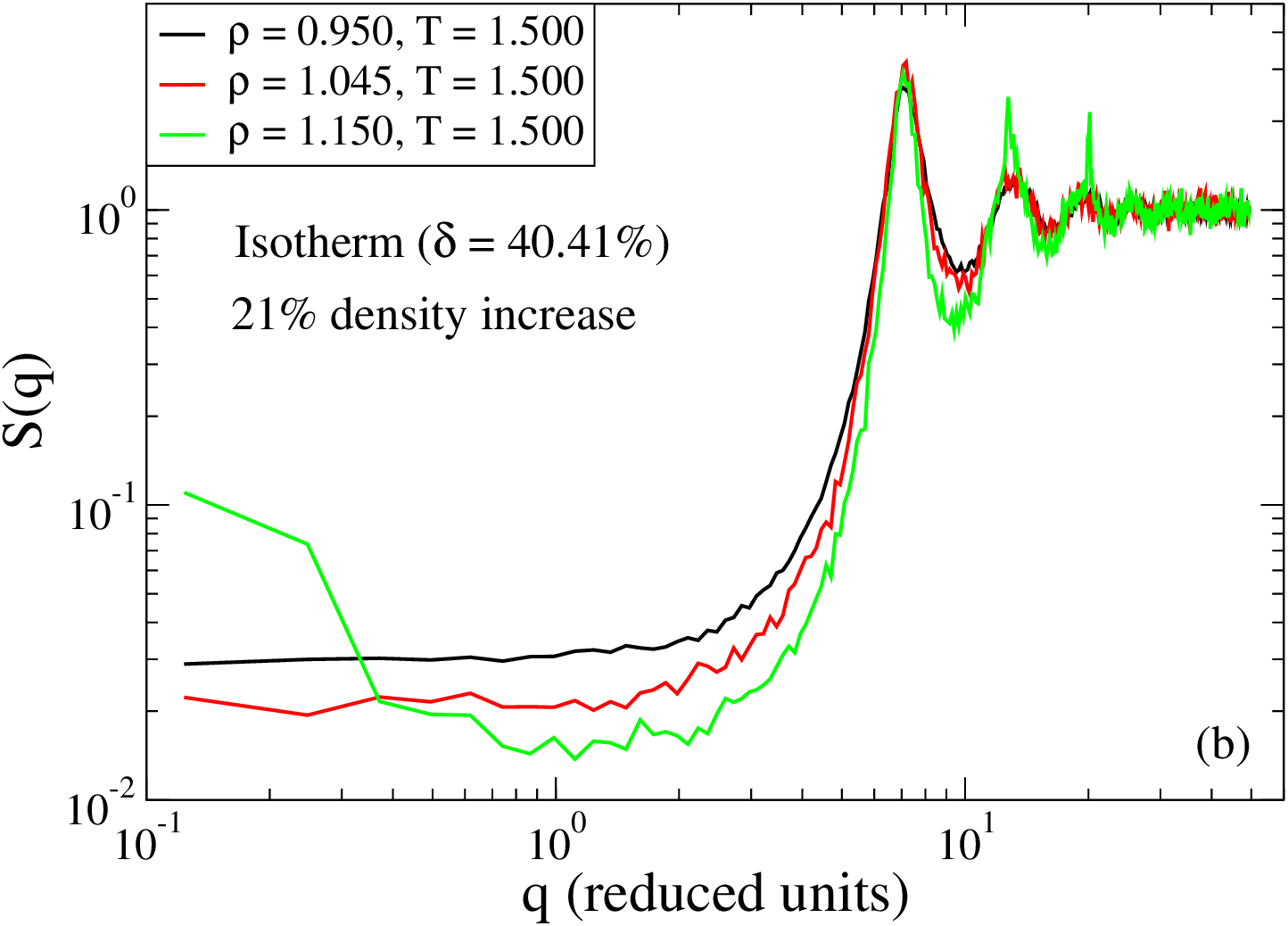}
  \caption{SSFs for the energy polydisperse LJ liquid of Figs. \ref{rdf3d1}-\ref{msd3d1}. (a) Along the isomorph. (b) Along the isotherm.}
  \label{sq3d1}
\end{figure}
Particle-resolved RDFs are presented in Fig. \ref{rdfp}. The low and mean $\epsilon_{i}$-value RDFs display good scaling along the isomorph (full and dotted curves), but the highest 
$\epsilon_{i}$-value RDFs exhibit small deviations (dashed curves). These deviations are attributed to the non-invariant phase separation.
\newline \newline
\begin{figure}[H]
  \centering
  \includegraphics[width=70mm]{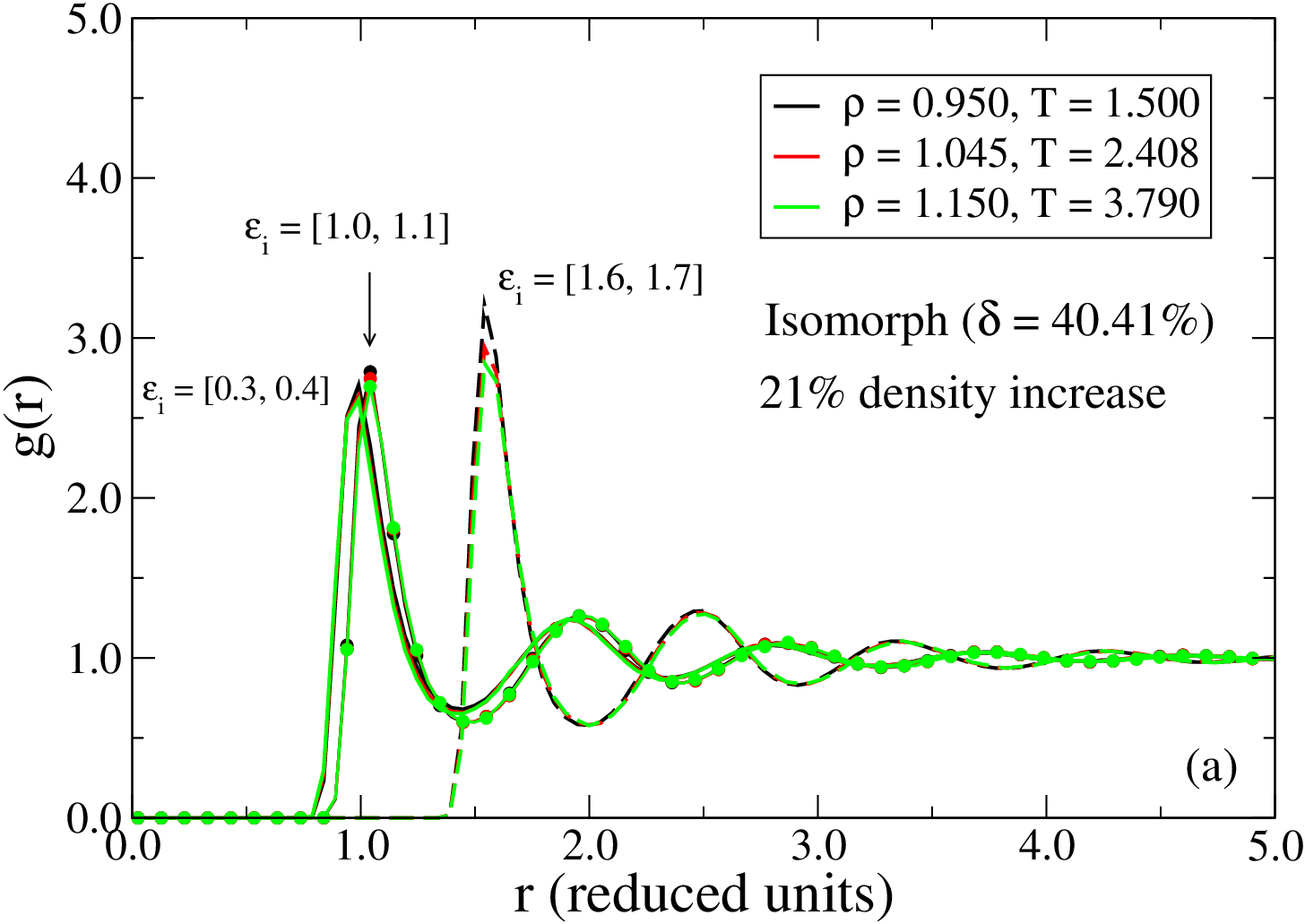}
\end{figure}
\begin{figure}[H]
  \centering
  \includegraphics[width=70mm]{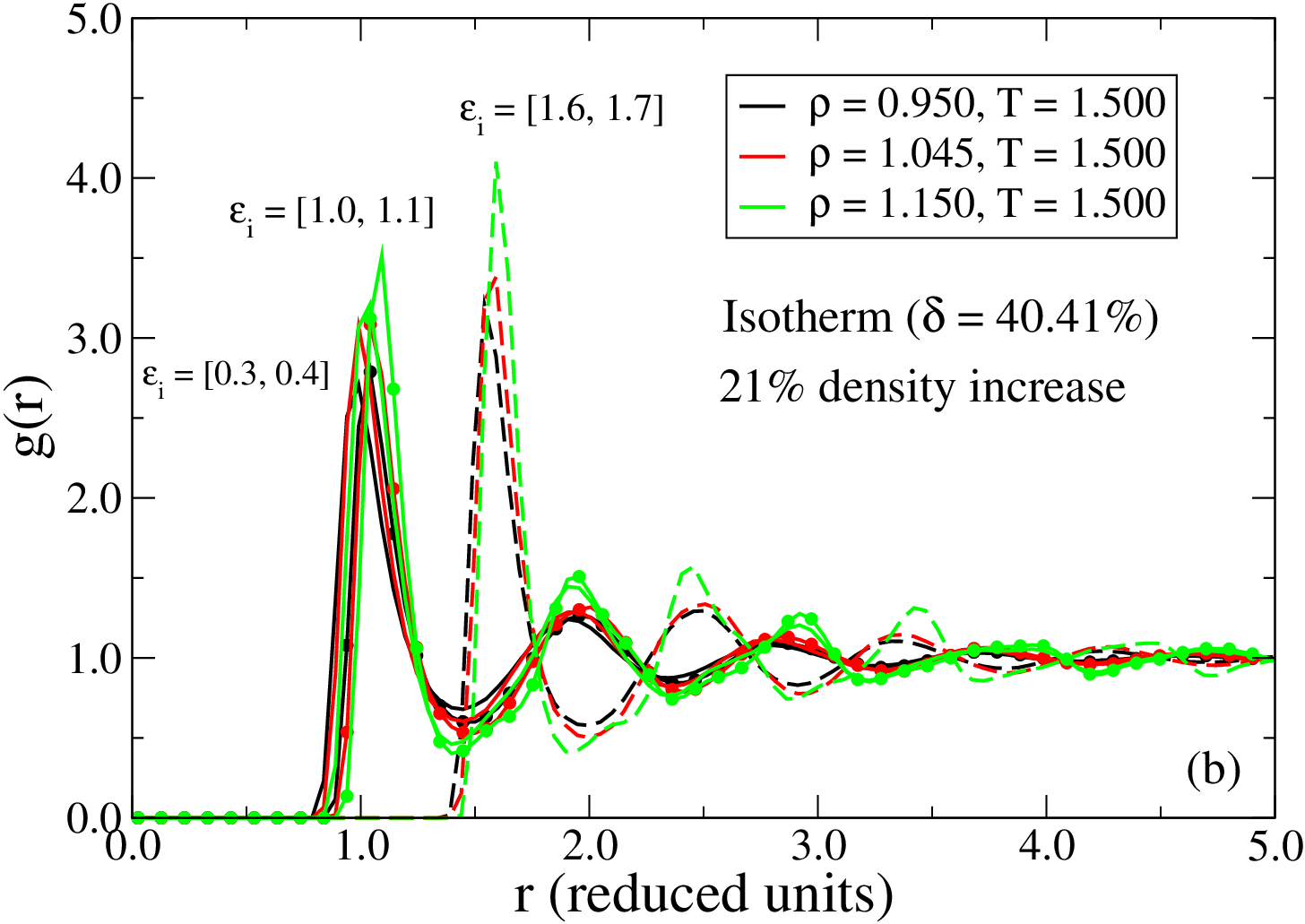}
  \caption{Particle-resolved RDFs for the energy polydisperse LJ liquid of Figs. \ref{rdf3d1}-\ref{sq3d1}. The full curves use particle energies in the range $\epsilon_{i}$ = [0.3, 0.4], 
    the dotted curves use $\epsilon_{i}$ = [1.0, 1.1], and the dashed curves use $\epsilon_{i}$ = [1.6, 1.7]. The dashed curves have 
    for clarity been shifted  with 0.5 in the $x$-direction. (a) Along the isomorph. (b) Along the isotherm.}
  \label{rdfp}
\end{figure}
The final isomorph considered in this study is an isomorph (R $\approx$ 0.94) for the modified energy polydisperse LJ system (Eqs. (\ref{newlj1}) and (\ref{newlj2})). Figure \ref{mod1} presents 
RDFs and MSDs along the isomorph where perfect scaling is observed.
\newline \newline
\begin{figure}[H]
  \centering
  \includegraphics[width=70mm]{new_iso_rdf}
\end{figure}
\begin{figure}[H]
  \centering
  \includegraphics[width=70mm]{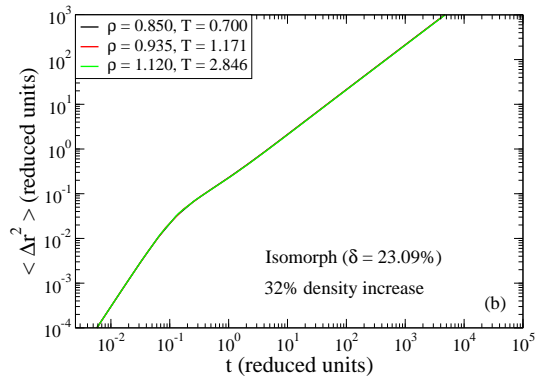}
  \caption{Invariance along an isomorph for the modified energy polydisperse LJ system (Eqs. (\ref{newlj1}) and (\ref{newlj2})) with $\delta$ = 23.09\%. (a) RDFs. (b) MSDs.}
  \label{mod1}
\end{figure}
Figure \ref{mod2} displays SSFs and particle-resolved RDFs along the same isomorph. The SSF and large $\epsilon_{i}$-particles are again noted to exhibit deviations along the isomorph 
for this moderately polydisperse system ($\delta$ = 23.09\%), and is ascribed to the phase separation as is evident from Fig. \ref{mod2}(a).
\newline \newline
\begin{figure}[H]
  \centering
  \includegraphics[width=70mm]{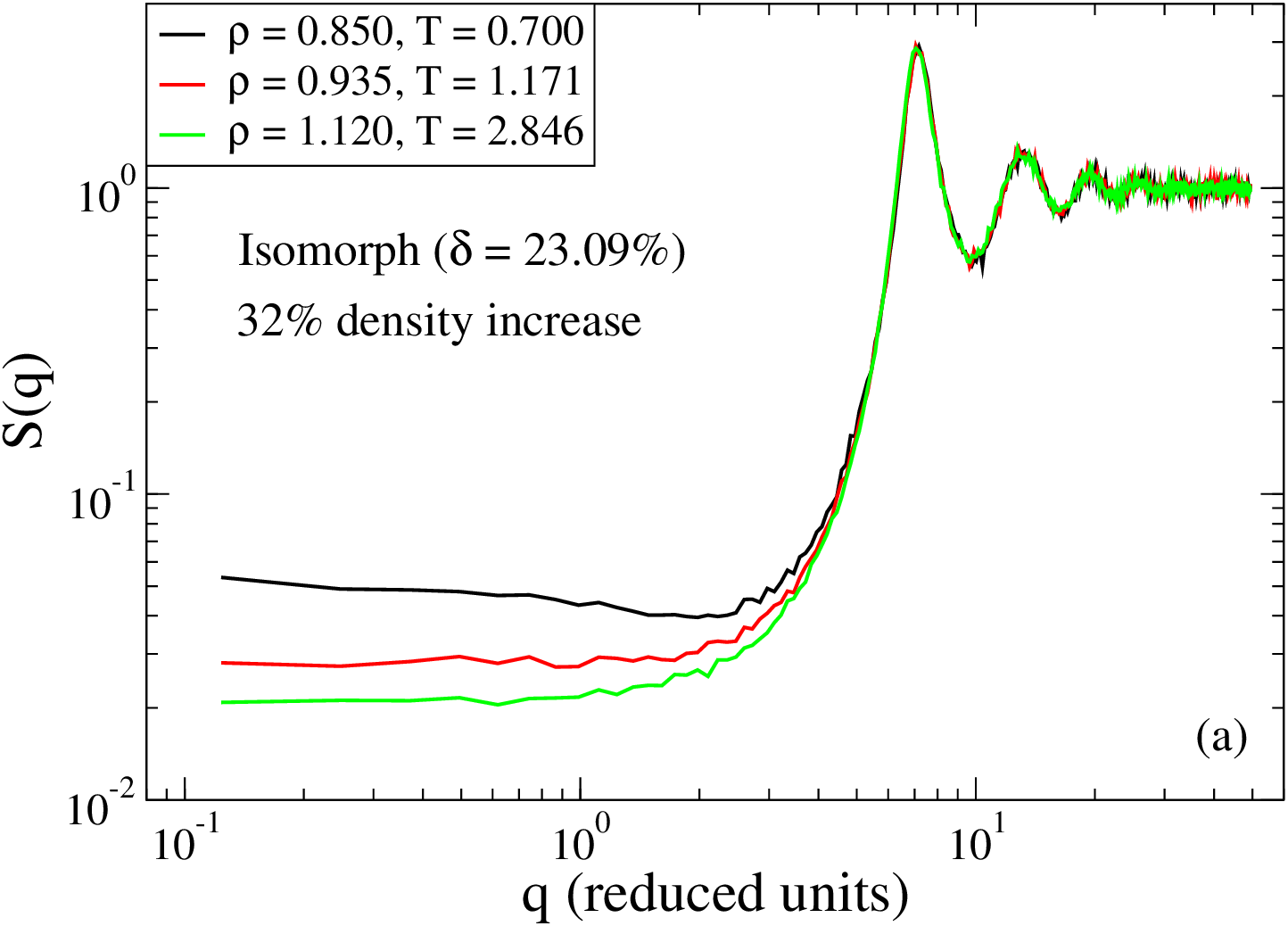}
\end{figure}
\begin{figure}[H]
  \centering
  \includegraphics[width=70mm]{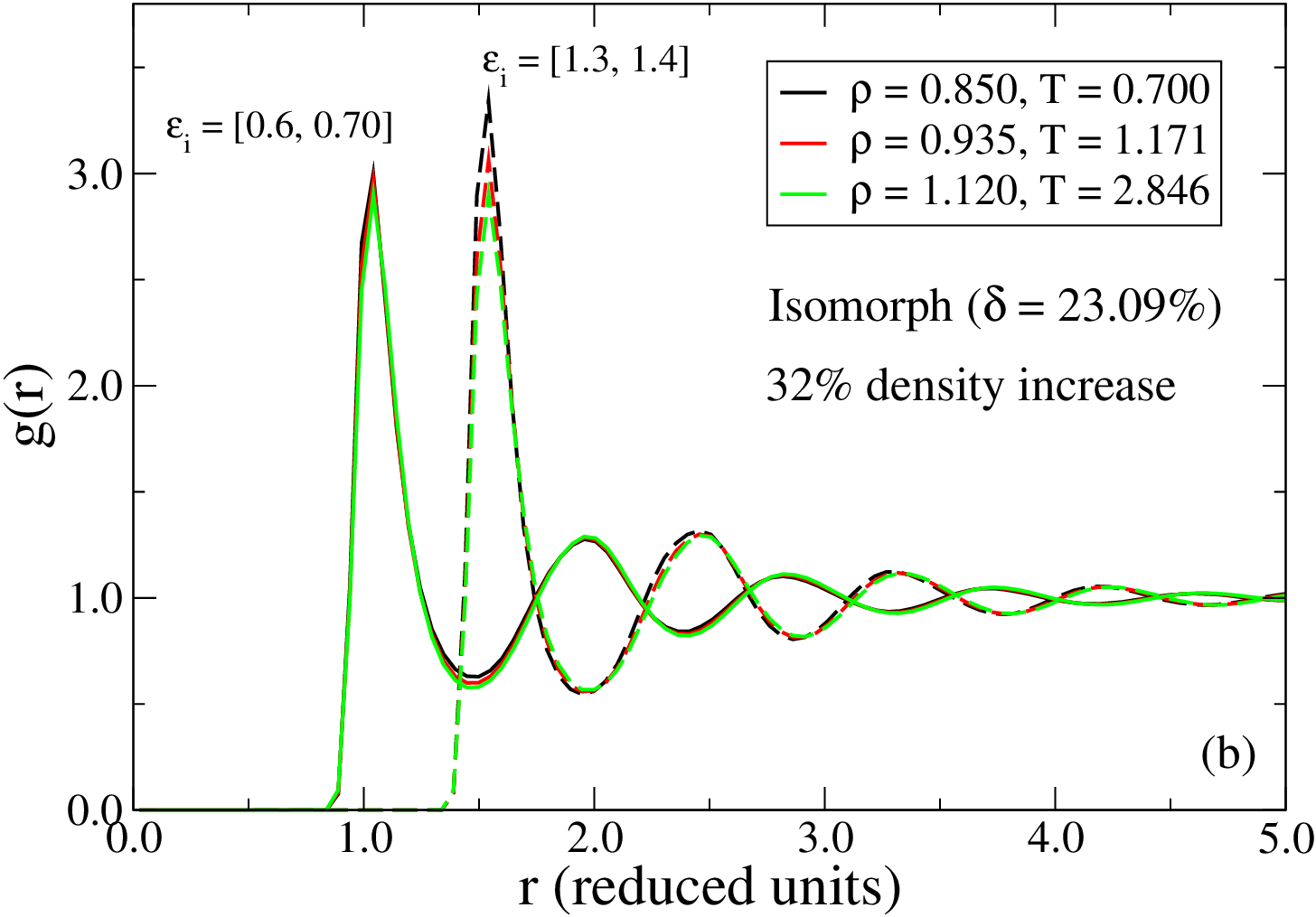}
  \caption{Invariance along the isomorph of Fig. \ref{mod1} for the modified energy polydisperse LJ system with $\delta$ = 23.09\%. (a) SSFs. (b) Particle-resolved RDFs. The full curves use 
    particle energies in the range $\epsilon_{i}$ = [0.6, 0.7], and the dashed curves use $\epsilon_{i}$ = [1.3, 1.4]. The dashed curves 
    have for clarity been shifted with 0.5 in the $x$-direction.}
  \label{mod2}
\end{figure}

\section{Conclusion and outlook}\label{con}

Polydisperse fluids are interesting to theoreticians and 
experimentalists alike owing to the perplexing phenomena these fluids display. In the companion paper\cite{ingebrigtsen2015} we studied the effect of size polydispersity 
on the nature of LJ liquids. The paper showed that even highly size polydisperse LJ liquids are RS liquids\cite{paper1}. Furthermore, it was shown that isomorphs extend 
readily to size polydisperse LJ liquids.

We considered in this paper the effect of energy polydispersity on the nature of LJ liquids. 
It was shown that highly energy polydisperse LJ liquids are also RS liquids. Particle segregation, however, leads to the loss of   
strong virial-potential energy correlation as a function of polydispersity. Depending on the investigated state point, this effect influences the strong virial-potential energy correlation to the extent 
that the liquid turns into a non-RS liquid. 

This decrease of the strong virial-potential energy correlation is a consequence of the fact that energy polydispersity leads to a serious modification of the phase behavior 
(see Fig. \ref{sep}). In comparison, for size polydisperse systems, the effects on the phase behavior are much weaker\cite{ingebrigtsen2015}. Isomorphs of energy polydisperse LJ liquids were also studied and shown to be a good approximation. Nevertheless, particle-resolved quantities showed larger deviations than the mean quantities along the isomorph due to the phase separation. This observation may thus affect particle-resolved scaling relations in polydisperse fluids\cite{pond2011}.

Motivation for the current study is derived from the fact that experimental 
polydisperse liquids are often dispersed in additional variables besides their size. As an example: Colloidal fluids can have 
a distribution in their size as well as their charge. Energy polydisperse liquids are less experimentally realizable than size polydisperse liquids. We 
believe, however, that the insight gained from this kind of study is imperative for understanding the effect of polydispersity on the nature of LJ liquids. 

Size and energy polydisperse LJ liquids are both RS liquids, however, the extent to which 
depends strongly on the type of polydispersity. Additional studies of polydispersity in relation to strong virial-potential energy correlation are thus needed 
to fully understand the effect of polydispersity on more complex RS liquids. The Yukawa potential was recently 
identified as RS in large parts of its phase diagram\cite{veldhorst2015}. It would be interesting to study a polydisperse version of this system and also of high relevance to experiments. 
We plan to address these issues in a later publication, and also welcome other studies of polydispersity and RS liquids.

\acknowledgments

This work was partially supported by Grants-in-Aid for Scientific Research (S) (21224011) and Specially Promoted Research (25000002) from the Japan Society of the Promotion of Science (JSPS).  T.S.I. acknowledges support from a JSPS Postdoctoral Fellowship.


\begin{thebibliography}{65}
\expandafter\ifx\csname natexlab\endcsname\relax\def\natexlab#1{#1}\fi
\expandafter\ifx\csname bibnamefont\endcsname\relax
  \def\bibnamefont#1{#1}\fi
\expandafter\ifx\csname bibfnamefont\endcsname\relax
  \def\bibfnamefont#1{#1}\fi
\expandafter\ifx\csname citenamefont\endcsname\relax
  \def\citenamefont#1{#1}\fi
\expandafter\ifx\csname url\endcsname\relax
  \def\url#1{\texttt{#1}}\fi
\expandafter\ifx\csname urlprefix\endcsname\relax\def\urlprefix{URL }\fi
\providecommand{\bibinfo}[2]{#2}
\providecommand{\eprint}[2][]{\url{#2}}

\bibitem[{\citenamefont{Bagchi}(2012)}]{bagchi2012}
\bibinfo{author}{\bibfnamefont{B.}~\bibnamefont{Bagchi}},
  \emph{\bibinfo{title}{Molecular Relaxation in Liquids}}
  (\bibinfo{publisher}{Oxford University Press: New York},
  \bibinfo{year}{2012}).

\bibitem[{\citenamefont{Wolynes and Lubchenko}(2012)}]{wolynes2012}
\bibinfo{editor}{\bibfnamefont{P.~G.} \bibnamefont{Wolynes}} \bibnamefont{and}
  \bibinfo{editor}{\bibfnamefont{V.}~\bibnamefont{Lubchenko}}, eds.,
  \emph{\bibinfo{title}{Structural Glasses and Supercooled Liquids: Theory,
  Experiment, and Applications}} (\bibinfo{publisher}{John Wiley and Sons,
  Inc.: New Jersey}, \bibinfo{year}{2012}).

\bibitem[{\citenamefont{Hansen et~al.}(2013)\citenamefont{Hansen, Lemarchand,
  Nielsen, Dyre, and Schr{\o}der}}]{hansen2013bit}
\bibinfo{author}{\bibfnamefont{J.~S.} \bibnamefont{Hansen}},
  \bibinfo{author}{\bibfnamefont{C.~A.} \bibnamefont{Lemarchand}},
  \bibinfo{author}{\bibfnamefont{E.}~\bibnamefont{Nielsen}},
  \bibinfo{author}{\bibfnamefont{J.~C.} \bibnamefont{Dyre}}, \bibnamefont{and}
  \bibinfo{author}{\bibfnamefont{T.}~\bibnamefont{Schr{\o}der}},
  \bibinfo{journal}{J. Chem. Phys.} \textbf{\bibinfo{volume}{138}},
  \bibinfo{pages}{094508} (\bibinfo{year}{2013}).

\bibitem[{\citenamefont{Dickinson}(1978)}]{dickinson1978}
\bibinfo{author}{\bibfnamefont{E.}~\bibnamefont{Dickinson}},
  \bibinfo{journal}{Chem. Phys. Lett.} \textbf{\bibinfo{volume}{57}},
  \bibinfo{pages}{148} (\bibinfo{year}{1978}).

\bibitem[{\citenamefont{Blum and Stell}(1979)}]{blum1979}
\bibinfo{author}{\bibfnamefont{L.}~\bibnamefont{Blum}} \bibnamefont{and}
  \bibinfo{author}{\bibfnamefont{G.}~\bibnamefont{Stell}}, \bibinfo{journal}{J.
  Chem. Phys.} \textbf{\bibinfo{volume}{71}}, \bibinfo{pages}{42}
  (\bibinfo{year}{1979}).

\bibitem[{\citenamefont{Salacuse and Stell}(1982)}]{salacuse1982}
\bibinfo{author}{\bibfnamefont{J.~J.} \bibnamefont{Salacuse}} \bibnamefont{and}
  \bibinfo{author}{\bibfnamefont{G.}~\bibnamefont{Stell}}, \bibinfo{journal}{J.
  Chem. Phys.} \textbf{\bibinfo{volume}{77}}, \bibinfo{pages}{3714}
  (\bibinfo{year}{1982}).

\bibitem[{\citenamefont{Ginoza and Yasutomi}(1997)}]{ginoza1997}
\bibinfo{author}{\bibfnamefont{M.}~\bibnamefont{Ginoza}} \bibnamefont{and}
  \bibinfo{author}{\bibfnamefont{M.}~\bibnamefont{Yasutomi}},
  \bibinfo{journal}{Mol. Phys.} \textbf{\bibinfo{volume}{91}},
  \bibinfo{pages}{59} (\bibinfo{year}{1997}).

\bibitem[{\citenamefont{Sollich}(2002)}]{sollich2002}
\bibinfo{author}{\bibfnamefont{P.}~\bibnamefont{Sollich}}, \bibinfo{journal}{J.
  Phys.: Condens. Matter} \textbf{\bibinfo{volume}{14}}, \bibinfo{pages}{R79}
  (\bibinfo{year}{2002}).

\bibitem[{\citenamefont{Fasolo and Sollich}(2003)}]{fasolo2003}
\bibinfo{author}{\bibfnamefont{M.}~\bibnamefont{Fasolo}} \bibnamefont{and}
  \bibinfo{author}{\bibfnamefont{P.}~\bibnamefont{Sollich}},
  \bibinfo{journal}{Phys. Rev. Lett.} \textbf{\bibinfo{volume}{91}},
  \bibinfo{pages}{068301} (\bibinfo{year}{2003}).

\bibitem[{\citenamefont{Jacobs and Frenkel}(2013)}]{jacobs2013}
\bibinfo{author}{\bibfnamefont{W.~M.} \bibnamefont{Jacobs}} \bibnamefont{and}
  \bibinfo{author}{\bibfnamefont{D.}~\bibnamefont{Frenkel}},
  \bibinfo{journal}{J. Chem. Phys.} \textbf{\bibinfo{volume}{139}},
  \bibinfo{pages}{024108} (\bibinfo{year}{2013}).

\bibitem[{\citenamefont{Frenkel et~al.}(1986)\citenamefont{Frenkel, Vos,
  de~Kruif, and Vrij}}]{frenkel1986}
\bibinfo{author}{\bibfnamefont{D.}~\bibnamefont{Frenkel}},
  \bibinfo{author}{\bibfnamefont{R.~J.} \bibnamefont{Vos}},
  \bibinfo{author}{\bibfnamefont{C.~G.} \bibnamefont{de~Kruif}},
  \bibnamefont{and} \bibinfo{author}{\bibfnamefont{A.}~\bibnamefont{Vrij}},
  \bibinfo{journal}{J. Chem. Phys.} \textbf{\bibinfo{volume}{84}},
  \bibinfo{pages}{4625} (\bibinfo{year}{1986}).

\bibitem[{\citenamefont{Kofke and Glandt}(1986)}]{kofke1986}
\bibinfo{author}{\bibfnamefont{D.~A.} \bibnamefont{Kofke}} \bibnamefont{and}
  \bibinfo{author}{\bibfnamefont{E.~D.} \bibnamefont{Glandt}},
  \bibinfo{journal}{FFE} \textbf{\bibinfo{volume}{29}}, \bibinfo{pages}{327}
  (\bibinfo{year}{1986}).

\bibitem[{\citenamefont{Wilding and Sollich}(2005)}]{wilding2005}
\bibinfo{author}{\bibfnamefont{N.~B.} \bibnamefont{Wilding}} \bibnamefont{and}
  \bibinfo{author}{\bibfnamefont{P.}~\bibnamefont{Sollich}},
  \bibinfo{journal}{J. Phys.: Condens. Matter} \textbf{\bibinfo{volume}{17}},
  \bibinfo{pages}{S3245} (\bibinfo{year}{2005}).

\bibitem[{\citenamefont{Wilding et~al.}(2005)\citenamefont{Wilding, Sollich,
  and Fasolo}}]{wilding2005b}
\bibinfo{author}{\bibfnamefont{N.~B.} \bibnamefont{Wilding}},
  \bibinfo{author}{\bibfnamefont{P.}~\bibnamefont{Sollich}}, \bibnamefont{and}
  \bibinfo{author}{\bibfnamefont{M.}~\bibnamefont{Fasolo}},
  \bibinfo{journal}{Phys. Rev. Lett.} \textbf{\bibinfo{volume}{95}},
  \bibinfo{pages}{155701} (\bibinfo{year}{2005}).

\bibitem[{\citenamefont{Wilding et~al.}(2006)\citenamefont{Wilding, Sollich,
  Fasolo, and Buzzacchi}}]{wilding2006}
\bibinfo{author}{\bibfnamefont{N.~B.} \bibnamefont{Wilding}},
  \bibinfo{author}{\bibfnamefont{P.}~\bibnamefont{Sollich}},
  \bibinfo{author}{\bibfnamefont{M.}~\bibnamefont{Fasolo}}, \bibnamefont{and}
  \bibinfo{author}{\bibfnamefont{M.}~\bibnamefont{Buzzacchi}},
  \bibinfo{journal}{J. Chem. Phys.} \textbf{\bibinfo{volume}{125}},
  \bibinfo{pages}{014908} (\bibinfo{year}{2006}).

\bibitem[{\citenamefont{Kawasaki et~al.}(2007)\citenamefont{Kawasaki, Araki,
  and Tanaka}}]{kawasaki2007}
\bibinfo{author}{\bibfnamefont{T.}~\bibnamefont{Kawasaki}},
  \bibinfo{author}{\bibfnamefont{T.}~\bibnamefont{Araki}}, \bibnamefont{and}
  \bibinfo{author}{\bibfnamefont{H.}~\bibnamefont{Tanaka}},
  \bibinfo{journal}{Phys. Rev. Lett.} \textbf{\bibinfo{volume}{99}},
  \bibinfo{pages}{215701} (\bibinfo{year}{2007}).

\bibitem[{\citenamefont{Sarkar et~al.}(2013)\citenamefont{Sarkar, Biswas,
  Santra, and Bagchi}}]{sarkar2013}
\bibinfo{author}{\bibfnamefont{S.}~\bibnamefont{Sarkar}},
  \bibinfo{author}{\bibfnamefont{R.}~\bibnamefont{Biswas}},
  \bibinfo{author}{\bibfnamefont{M.}~\bibnamefont{Santra}}, \bibnamefont{and}
  \bibinfo{author}{\bibfnamefont{B.}~\bibnamefont{Bagchi}},
  \bibinfo{journal}{Phys. Rev. E} \textbf{\bibinfo{volume}{88}},
  \bibinfo{pages}{022104} (\bibinfo{year}{2013}).

\bibitem[{\citenamefont{Sarkar et~al.}(2014)\citenamefont{Sarkar, Biswas, Ray,
  and Bagchi}}]{sarkar2014}
\bibinfo{author}{\bibfnamefont{S.}~\bibnamefont{Sarkar}},
  \bibinfo{author}{\bibfnamefont{R.}~\bibnamefont{Biswas}},
  \bibinfo{author}{\bibfnamefont{P.~P.} \bibnamefont{Ray}}, \bibnamefont{and}
  \bibinfo{author}{\bibfnamefont{B.}~\bibnamefont{Bagchi}},
  \bibinfo{journal}{arXiv} p. \bibinfo{pages}{1402.6879}
  (\bibinfo{year}{2014}).

\bibitem[{\citenamefont{Koningsveld and Kleintjens}(1971)}]{koningsveld1971}
\bibinfo{author}{\bibfnamefont{R.}~\bibnamefont{Koningsveld}} \bibnamefont{and}
  \bibinfo{author}{\bibfnamefont{L.~A.} \bibnamefont{Kleintjens}},
  \bibinfo{journal}{Macromolecules} \textbf{\bibinfo{volume}{4}},
  \bibinfo{pages}{637} (\bibinfo{year}{1971}).

\bibitem[{\citenamefont{Cowell and Vincent}(1982)}]{cowell1982}
\bibinfo{author}{\bibfnamefont{C.}~\bibnamefont{Cowell}} \bibnamefont{and}
  \bibinfo{author}{\bibfnamefont{B.}~\bibnamefont{Vincent}},
  \bibinfo{journal}{J. Colloid Interface Sci.} \textbf{\bibinfo{volume}{87}},
  \bibinfo{pages}{518} (\bibinfo{year}{1982}).

\bibitem[{\citenamefont{Weeks et~al.}(2000)\citenamefont{Weeks, Crocker,
  Levitt, Schofield, and Weitz}}]{weeks2000}
\bibinfo{author}{\bibfnamefont{E.~R.} \bibnamefont{Weeks}},
  \bibinfo{author}{\bibfnamefont{J.~C.} \bibnamefont{Crocker}},
  \bibinfo{author}{\bibfnamefont{A.~C.} \bibnamefont{Levitt}},
  \bibinfo{author}{\bibfnamefont{A.}~\bibnamefont{Schofield}},
  \bibnamefont{and} \bibinfo{author}{\bibfnamefont{D.~A.} \bibnamefont{Weitz}},
  \bibinfo{journal}{Science} \textbf{\bibinfo{volume}{287}},
  \bibinfo{pages}{627} (\bibinfo{year}{2000}).

\bibitem[{\citenamefont{Auer and Frenkel}(2001)}]{auer2001}
\bibinfo{author}{\bibfnamefont{S.}~\bibnamefont{Auer}} \bibnamefont{and}
  \bibinfo{author}{\bibfnamefont{D.}~\bibnamefont{Frenkel}},
  \bibinfo{journal}{Nature} \textbf{\bibinfo{volume}{413}},
  \bibinfo{pages}{711} (\bibinfo{year}{2001}).

\bibitem[{\citenamefont{Watanabe and Tanaka}(2008)}]{watanabe2008}
\bibinfo{author}{\bibfnamefont{K.}~\bibnamefont{Watanabe}} \bibnamefont{and}
  \bibinfo{author}{\bibfnamefont{H.}~\bibnamefont{Tanaka}},
  \bibinfo{journal}{Phys. Rev. Lett.} \textbf{\bibinfo{volume}{100}},
  \bibinfo{pages}{158002} (\bibinfo{year}{2008}).

\bibitem[{\citenamefont{Sacanna et~al.}(2013)\citenamefont{Sacanna, Korpics,
  Rodriguez, Col{\'o}n-Mel{\'e}ndez, Kim, Pine, and Yi}}]{sacanna2013}
\bibinfo{author}{\bibfnamefont{S.}~\bibnamefont{Sacanna}},
  \bibinfo{author}{\bibfnamefont{M.}~\bibnamefont{Korpics}},
  \bibinfo{author}{\bibfnamefont{K.}~\bibnamefont{Rodriguez}},
  \bibinfo{author}{\bibfnamefont{L.}~\bibnamefont{Col{\'o}n-Mel{\'e}ndez}},
  \bibinfo{author}{\bibfnamefont{S.-H.} \bibnamefont{Kim}},
  \bibinfo{author}{\bibfnamefont{D.~J.} \bibnamefont{Pine}}, \bibnamefont{and}
  \bibinfo{author}{\bibfnamefont{G.-R.} \bibnamefont{Yi}},
  \bibinfo{journal}{Nat. Commun.} \textbf{\bibinfo{volume}{4}},
  \bibinfo{pages}{1688} (\bibinfo{year}{2013}).

\bibitem[{\citenamefont{Palberg}(2014)}]{palberg2014}
\bibinfo{author}{\bibfnamefont{T.}~\bibnamefont{Palberg}}, \bibinfo{journal}{J.
  Phys.: Condens. Matter} \textbf{\bibinfo{volume}{26}},
  \bibinfo{pages}{333101} (\bibinfo{year}{2014}).

\bibitem[{\citenamefont{Leocmach and Tanaka}(2012)}]{leocmach2012}
\bibinfo{author}{\bibfnamefont{M.}~\bibnamefont{Leocmach}} \bibnamefont{and}
  \bibinfo{author}{\bibfnamefont{H.}~\bibnamefont{Tanaka}},
  \bibinfo{journal}{Nat. Commun.} \textbf{\bibinfo{volume}{3}},
  \bibinfo{pages}{974} (\bibinfo{year}{2012}).

\bibitem[{\citenamefont{Tanaka}(2012)}]{tanaka_review}
\bibinfo{author}{\bibfnamefont{H.}~\bibnamefont{Tanaka}},
  \bibinfo{journal}{Eur. Phys. J E} \textbf{\bibinfo{volume}{35}},
  \bibinfo{pages}{113} (\bibinfo{year}{2012}).

\bibitem[{\citenamefont{Leocmach et~al.}(2013)\citenamefont{Leocmach, Russo,
  and Tanaka}}]{mathieu_russo_tanaka}
\bibinfo{author}{\bibfnamefont{M.}~\bibnamefont{Leocmach}},
  \bibinfo{author}{\bibfnamefont{J.}~\bibnamefont{Russo}}, \bibnamefont{and}
  \bibinfo{author}{\bibfnamefont{H.}~\bibnamefont{Tanaka}},
  \bibinfo{journal}{J. Chem. Phys.} \textbf{\bibinfo{volume}{138}},
  \bibinfo{eid}{12A536} (\bibinfo{year}{2013}).

\bibitem[{\citenamefont{Evans}(1999)}]{evans1999}
\bibinfo{author}{\bibfnamefont{R.~M.~L.} \bibnamefont{Evans}},
  \bibinfo{journal}{Phys. Rev. E} \textbf{\bibinfo{volume}{59}},
  \bibinfo{pages}{3192} (\bibinfo{year}{1999}).

\bibitem[{\citenamefont{Ingebrigtsen and Tanaka}(2015)}]{ingebrigtsen2015}
\bibinfo{author}{\bibfnamefont{T.~S.} \bibnamefont{Ingebrigtsen}}
  \bibnamefont{and} \bibinfo{author}{\bibfnamefont{H.}~\bibnamefont{Tanaka}},
  \bibinfo{journal}{J. Phys. Chem. B} \textbf{\bibinfo{volume}{119}},
  \bibinfo{pages}{11052} (\bibinfo{year}{2015}).

\bibitem[{\citenamefont{Pedersen et~al.}(2008)\citenamefont{Pedersen, Bailey,
  Schr{\o}der, and Dyre}}]{paper0}
\bibinfo{author}{\bibfnamefont{U.~R.} \bibnamefont{Pedersen}},
  \bibinfo{author}{\bibfnamefont{N.~P.} \bibnamefont{Bailey}},
  \bibinfo{author}{\bibfnamefont{T.~B.} \bibnamefont{Schr{\o}der}},
  \bibnamefont{and} \bibinfo{author}{\bibfnamefont{J.~C.} \bibnamefont{Dyre}},
  \bibinfo{journal}{Phys. Rev. Lett.} \textbf{\bibinfo{volume}{100}},
  \bibinfo{pages}{015701} (\bibinfo{year}{2008}).

\bibitem[{\citenamefont{Bailey et~al.}(2008{\natexlab{a}})\citenamefont{Bailey,
  Pedersen, Gnan, Schr{\o}der, and Dyre}}]{paper1}
\bibinfo{author}{\bibfnamefont{N.~P.} \bibnamefont{Bailey}},
  \bibinfo{author}{\bibfnamefont{U.~R.} \bibnamefont{Pedersen}},
  \bibinfo{author}{\bibfnamefont{N.}~\bibnamefont{Gnan}},
  \bibinfo{author}{\bibfnamefont{T.~B.} \bibnamefont{Schr{\o}der}},
  \bibnamefont{and} \bibinfo{author}{\bibfnamefont{J.~C.} \bibnamefont{Dyre}},
  \bibinfo{journal}{J. Chem. Phys.} \textbf{\bibinfo{volume}{129}},
  \bibinfo{pages}{184507} (\bibinfo{year}{2008}{\natexlab{a}}).

\bibitem[{\citenamefont{Bailey et~al.}(2008{\natexlab{b}})\citenamefont{Bailey,
  Pedersen, Gnan, Schr{\o}der, and Dyre}}]{paper2}
\bibinfo{author}{\bibfnamefont{N.~P.} \bibnamefont{Bailey}},
  \bibinfo{author}{\bibfnamefont{U.~R.} \bibnamefont{Pedersen}},
  \bibinfo{author}{\bibfnamefont{N.}~\bibnamefont{Gnan}},
  \bibinfo{author}{\bibfnamefont{T.~B.} \bibnamefont{Schr{\o}der}},
  \bibnamefont{and} \bibinfo{author}{\bibfnamefont{J.~C.} \bibnamefont{Dyre}},
  \bibinfo{journal}{J. Chem. Phys.} \textbf{\bibinfo{volume}{129}},
  \bibinfo{pages}{184508} (\bibinfo{year}{2008}{\natexlab{b}}).

\bibitem[{\citenamefont{Schr{\o}der
  et~al.}(2009{\natexlab{a}})\citenamefont{Schr{\o}der, Bailey, Pedersen, Gnan,
  and Dyre}}]{paper3}
\bibinfo{author}{\bibfnamefont{T.~B.} \bibnamefont{Schr{\o}der}},
  \bibinfo{author}{\bibfnamefont{N.~P.} \bibnamefont{Bailey}},
  \bibinfo{author}{\bibfnamefont{U.~R.} \bibnamefont{Pedersen}},
  \bibinfo{author}{\bibfnamefont{N.}~\bibnamefont{Gnan}}, \bibnamefont{and}
  \bibinfo{author}{\bibfnamefont{J.~C.} \bibnamefont{Dyre}},
  \bibinfo{journal}{J. Chem. Phys.} \textbf{\bibinfo{volume}{131}},
  \bibinfo{pages}{234503} (\bibinfo{year}{2009}{\natexlab{a}}).

\bibitem[{\citenamefont{Gnan et~al.}(2009)\citenamefont{Gnan, Schr{\o}der,
  Pedersen, Bailey, and Dyre}}]{paper4}
\bibinfo{author}{\bibfnamefont{N.}~\bibnamefont{Gnan}},
  \bibinfo{author}{\bibfnamefont{T.~B.} \bibnamefont{Schr{\o}der}},
  \bibinfo{author}{\bibfnamefont{U.~R.} \bibnamefont{Pedersen}},
  \bibinfo{author}{\bibfnamefont{N.~P.} \bibnamefont{Bailey}},
  \bibnamefont{and} \bibinfo{author}{\bibfnamefont{J.~C.} \bibnamefont{Dyre}},
  \bibinfo{journal}{J. Chem. Phys.} \textbf{\bibinfo{volume}{131}},
  \bibinfo{pages}{234504} (\bibinfo{year}{2009}).

\bibitem[{\citenamefont{Schr{\o}der et~al.}(2011)\citenamefont{Schr{\o}der,
  Gnan, Pedersen, Bailey, and Dyre}}]{paper5}
\bibinfo{author}{\bibfnamefont{T.~B.} \bibnamefont{Schr{\o}der}},
  \bibinfo{author}{\bibfnamefont{N.}~\bibnamefont{Gnan}},
  \bibinfo{author}{\bibfnamefont{U.~R.} \bibnamefont{Pedersen}},
  \bibinfo{author}{\bibfnamefont{N.~P.} \bibnamefont{Bailey}},
  \bibnamefont{and} \bibinfo{author}{\bibfnamefont{J.~C.} \bibnamefont{Dyre}},
  \bibinfo{journal}{J. Chem. Phys.} \textbf{\bibinfo{volume}{134}},
  \bibinfo{pages}{164505} (\bibinfo{year}{2011}).

\bibitem[{\citenamefont{Ingebrigtsen
  et~al.}(2012{\natexlab{a}})\citenamefont{Ingebrigtsen, Schr{\o}der, and
  Dyre}}]{prx}
\bibinfo{author}{\bibfnamefont{T.~S.} \bibnamefont{Ingebrigtsen}},
  \bibinfo{author}{\bibfnamefont{T.~B.} \bibnamefont{Schr{\o}der}},
  \bibnamefont{and} \bibinfo{author}{\bibfnamefont{J.~C.} \bibnamefont{Dyre}},
  \bibinfo{journal}{Phys. Rev. X} \textbf{\bibinfo{volume}{2}},
  \bibinfo{pages}{011011} (\bibinfo{year}{2012}{\natexlab{a}}).

\bibitem[{\citenamefont{Ingebrigtsen
  et~al.}(2012{\natexlab{b}})\citenamefont{Ingebrigtsen, Schr{\o}der, and
  Dyre}}]{moleculesisomorphs}
\bibinfo{author}{\bibfnamefont{T.~S.} \bibnamefont{Ingebrigtsen}},
  \bibinfo{author}{\bibfnamefont{T.~B.} \bibnamefont{Schr{\o}der}},
  \bibnamefont{and} \bibinfo{author}{\bibfnamefont{J.~C.} \bibnamefont{Dyre}},
  \bibinfo{journal}{J. Phys. Chem. B} \textbf{\bibinfo{volume}{116}},
  \bibinfo{pages}{1018} (\bibinfo{year}{2012}{\natexlab{b}}).

\bibitem[{\citenamefont{Largo and Wilding}(2006)}]{largo2006}
\bibinfo{author}{\bibfnamefont{J.}~\bibnamefont{Largo}} \bibnamefont{and}
  \bibinfo{author}{\bibfnamefont{N.~B.} \bibnamefont{Wilding}},
  \bibinfo{journal}{Phys. Rev. E} \textbf{\bibinfo{volume}{73}},
  \bibinfo{pages}{036115} (\bibinfo{year}{2006}).

\bibitem[{\citenamefont{Stapleton et~al.}(1988)\citenamefont{Stapleton,
  Tildesley, Sluckin, and N}}]{stapleton1988}
\bibinfo{author}{\bibfnamefont{M.~R.} \bibnamefont{Stapleton}},
  \bibinfo{author}{\bibfnamefont{D.~J.} \bibnamefont{Tildesley}},
  \bibinfo{author}{\bibfnamefont{T.~J.} \bibnamefont{Sluckin}},
  \bibnamefont{and} \bibinfo{author}{\bibfnamefont{Q.}~\bibnamefont{N}},
  \bibinfo{journal}{J. Phys. Chem.} \textbf{\bibinfo{volume}{92}},
  \bibinfo{pages}{4788} (\bibinfo{year}{1988}).

\bibitem[{\citenamefont{Shagolsem et~al.}(2015)\citenamefont{Shagolsem,
  Osmanovi{\'c}, Peleg, and Rabin}}]{shagolsem2015a}
\bibinfo{author}{\bibfnamefont{L.~S.} \bibnamefont{Shagolsem}},
  \bibinfo{author}{\bibfnamefont{D.}~\bibnamefont{Osmanovi{\'c}}},
  \bibinfo{author}{\bibfnamefont{O.}~\bibnamefont{Peleg}}, \bibnamefont{and}
  \bibinfo{author}{\bibfnamefont{Y.}~\bibnamefont{Rabin}}, \bibinfo{journal}{J.
  Chem. Phys.} \textbf{\bibinfo{volume}{142}}, \bibinfo{pages}{051104}
  (\bibinfo{year}{2015}).

\bibitem[{\citenamefont{Shagolsem and Rabin}(2015)}]{shagolsem2015b}
\bibinfo{author}{\bibfnamefont{L.~S.} \bibnamefont{Shagolsem}}
  \bibnamefont{and} \bibinfo{author}{\bibfnamefont{Y.}~\bibnamefont{Rabin}},
  \bibinfo{journal}{arXiv} p. \bibinfo{pages}{1508.05264v1}
  (\bibinfo{year}{2015}).

\bibitem[{\citenamefont{Osmanovi{\'c} and Rabin}(2016)}]{rabin2016}
\bibinfo{author}{\bibfnamefont{D.}~\bibnamefont{Osmanovi{\'c}}}
  \bibnamefont{and} \bibinfo{author}{\bibfnamefont{Y.}~\bibnamefont{Rabin}},
  \bibinfo{journal}{J. Stat. Phys.} \textbf{\bibinfo{volume}{162}},
  \bibinfo{pages}{186} (\bibinfo{year}{2016}).

\bibitem[{\citenamefont{Toxvaerd}(1991)}]{nvttoxvaerd}
\bibinfo{author}{\bibfnamefont{S.}~\bibnamefont{Toxvaerd}},
  \bibinfo{journal}{Mol. Phys.} \textbf{\bibinfo{volume}{72}},
  \bibinfo{pages}{159} (\bibinfo{year}{1991}).

\bibitem[{\citenamefont{Bailey et~al.}(2015)\citenamefont{Bailey, Ingebrigtsen,
  Hansen, Veldhorst, B{\o}hling, Lemarchand, Olsen, Bacher, Larsen, Dyre
  et~al.}}]{rumd}
\bibinfo{author}{\bibfnamefont{N.~P.} \bibnamefont{Bailey}},
  \bibinfo{author}{\bibfnamefont{T.~S.} \bibnamefont{Ingebrigtsen}},
  \bibinfo{author}{\bibfnamefont{J.~S.} \bibnamefont{Hansen}},
  \bibinfo{author}{\bibfnamefont{A.~A.} \bibnamefont{Veldhorst}},
  \bibinfo{author}{\bibfnamefont{L.}~\bibnamefont{B{\o}hling}},
  \bibinfo{author}{\bibfnamefont{C.~A.} \bibnamefont{Lemarchand}},
  \bibinfo{author}{\bibfnamefont{A.~E.} \bibnamefont{Olsen}},
  \bibinfo{author}{\bibfnamefont{A.~K.} \bibnamefont{Bacher}},
  \bibinfo{author}{\bibfnamefont{H.}~\bibnamefont{Larsen}},
  \bibinfo{author}{\bibfnamefont{J.~C.} \bibnamefont{Dyre}},
  \bibnamefont{et~al.}, \bibinfo{journal}{arXiv} p. \bibinfo{pages}{1506.05094}
  (\bibinfo{year}{2015}).

\bibitem[{\citenamefont{Allen and Tildesley}(1987)}]{tildesley}
\bibinfo{author}{\bibfnamefont{M.~P.} \bibnamefont{Allen}} \bibnamefont{and}
  \bibinfo{author}{\bibfnamefont{D.~J.} \bibnamefont{Tildesley}},
  \emph{\bibinfo{title}{Computer Simulation of Liquids}}
  (\bibinfo{publisher}{Oxford University Press: New York},
  \bibinfo{year}{1987}).

\bibitem[{\citenamefont{Dyre}(2014)}]{dyre2014}
\bibinfo{author}{\bibfnamefont{J.~C.} \bibnamefont{Dyre}}, \bibinfo{journal}{J.
  Phys. Chem. B} \textbf{\bibinfo{volume}{118}}, \bibinfo{pages}{10007}
  (\bibinfo{year}{2014}).

\bibitem[{\citenamefont{Albrechtsen et~al.}(2014)\citenamefont{Albrechtsen,
  Olsen, Pedersen, Schr{\o}der, and Dyre}}]{crystals}
\bibinfo{author}{\bibfnamefont{D.~E.} \bibnamefont{Albrechtsen}},
  \bibinfo{author}{\bibfnamefont{A.~E.} \bibnamefont{Olsen}},
  \bibinfo{author}{\bibfnamefont{U.~R.} \bibnamefont{Pedersen}},
  \bibinfo{author}{\bibfnamefont{T.~B.} \bibnamefont{Schr{\o}der}},
  \bibnamefont{and} \bibinfo{author}{\bibfnamefont{J.~C.} \bibnamefont{Dyre}},
  \bibinfo{journal}{Phys. Rev. B} \textbf{\bibinfo{volume}{90}},
  \bibinfo{pages}{094106} (\bibinfo{year}{2014}).

\bibitem[{\citenamefont{Coslovich and Roland}(2008)}]{coslovich1}
\bibinfo{author}{\bibfnamefont{D.}~\bibnamefont{Coslovich}} \bibnamefont{and}
  \bibinfo{author}{\bibfnamefont{C.~M.} \bibnamefont{Roland}},
  \bibinfo{journal}{J. Phys. Chem. B} \textbf{\bibinfo{volume}{112}},
  \bibinfo{pages}{1329} (\bibinfo{year}{2008}).

\bibitem[{\citenamefont{Coslovich and Roland}(2009)}]{coslovich2}
\bibinfo{author}{\bibfnamefont{D.}~\bibnamefont{Coslovich}} \bibnamefont{and}
  \bibinfo{author}{\bibfnamefont{C.~M.} \bibnamefont{Roland}},
  \bibinfo{journal}{J. Chem. Phys.} \textbf{\bibinfo{volume}{130}},
  \bibinfo{pages}{014508} (\bibinfo{year}{2009}).

\bibitem[{\citenamefont{Schr{\o}der
  et~al.}(2009{\natexlab{b}})\citenamefont{Schr{\o}der, Pedersen, Bailey,
  Toxvaerd, and Dyre}}]{moleculeshidden}
\bibinfo{author}{\bibfnamefont{T.~B.} \bibnamefont{Schr{\o}der}},
  \bibinfo{author}{\bibfnamefont{U.~R.} \bibnamefont{Pedersen}},
  \bibinfo{author}{\bibfnamefont{N.~P.} \bibnamefont{Bailey}},
  \bibinfo{author}{\bibfnamefont{S.}~\bibnamefont{Toxvaerd}}, \bibnamefont{and}
  \bibinfo{author}{\bibfnamefont{J.~C.} \bibnamefont{Dyre}},
  \bibinfo{journal}{Phys. Rev. E} \textbf{\bibinfo{volume}{80}},
  \bibinfo{pages}{041502} (\bibinfo{year}{2009}{\natexlab{b}}).

\bibitem[{\citenamefont{Kob and Andersen}(1995)}]{ka1}
\bibinfo{author}{\bibfnamefont{W.}~\bibnamefont{Kob}} \bibnamefont{and}
  \bibinfo{author}{\bibfnamefont{H.~C.} \bibnamefont{Andersen}},
  \bibinfo{journal}{Phys. Rev. E} \textbf{\bibinfo{volume}{51}},
  \bibinfo{pages}{4626} (\bibinfo{year}{1995}).

\bibitem[{\citenamefont{Lewis and Wahnstr{\"o}m}(1994)}]{otp1}
\bibinfo{author}{\bibfnamefont{L.~J.} \bibnamefont{Lewis}} \bibnamefont{and}
  \bibinfo{author}{\bibfnamefont{G.}~\bibnamefont{Wahnstr{\"o}m}},
  \bibinfo{journal}{J. Non-Cryst. Solids} \textbf{\bibinfo{volume}{172-174}},
  \bibinfo{pages}{69} (\bibinfo{year}{1994}).

\bibitem[{\citenamefont{Gundermann et~al.}(2011)\citenamefont{Gundermann,
  Pedersen, Hecksher, Bailey, Jakobsen, Christensen, Olsen, Schr{\o}der,
  Fragiadakis, Casalini et~al.}}]{gammagamma}
\bibinfo{author}{\bibfnamefont{D.}~\bibnamefont{Gundermann}},
  \bibinfo{author}{\bibfnamefont{U.~R.} \bibnamefont{Pedersen}},
  \bibinfo{author}{\bibfnamefont{T.}~\bibnamefont{Hecksher}},
  \bibinfo{author}{\bibfnamefont{N.~P.} \bibnamefont{Bailey}},
  \bibinfo{author}{\bibfnamefont{B.}~\bibnamefont{Jakobsen}},
  \bibinfo{author}{\bibfnamefont{T.}~\bibnamefont{Christensen}},
  \bibinfo{author}{\bibfnamefont{N.~B.} \bibnamefont{Olsen}},
  \bibinfo{author}{\bibfnamefont{T.~B.} \bibnamefont{Schr{\o}der}},
  \bibinfo{author}{\bibfnamefont{D.}~\bibnamefont{Fragiadakis}},
  \bibinfo{author}{\bibfnamefont{R.}~\bibnamefont{Casalini}},
  \bibnamefont{et~al.}, \bibinfo{journal}{Nat. Phys.}
  \textbf{\bibinfo{volume}{7}}, \bibinfo{pages}{816} (\bibinfo{year}{2011}).

\bibitem[{\citenamefont{Roed et~al.}(2013)\citenamefont{Roed, Gundermann, Dyre,
  and Niss}}]{roed2013}
\bibinfo{author}{\bibfnamefont{L.~A.} \bibnamefont{Roed}},
  \bibinfo{author}{\bibfnamefont{D.}~\bibnamefont{Gundermann}},
  \bibinfo{author}{\bibfnamefont{J.~C.} \bibnamefont{Dyre}}, \bibnamefont{and}
  \bibinfo{author}{\bibfnamefont{K.}~\bibnamefont{Niss}}, \bibinfo{journal}{J.
  Chem. Phys.} \textbf{\bibinfo{volume}{139}}, \bibinfo{pages}{101101}
  (\bibinfo{year}{2013}).

\bibitem[{\citenamefont{Xiao et~al.}(2015)\citenamefont{Xiao, Tofteskov,
  Christensen, Dyre, and Niss}}]{wence2014}
\bibinfo{author}{\bibfnamefont{W.}~\bibnamefont{Xiao}},
  \bibinfo{author}{\bibfnamefont{J.}~\bibnamefont{Tofteskov}},
  \bibinfo{author}{\bibfnamefont{T.~V.} \bibnamefont{Christensen}},
  \bibinfo{author}{\bibfnamefont{J.~C.} \bibnamefont{Dyre}}, \bibnamefont{and}
  \bibinfo{author}{\bibfnamefont{K.}~\bibnamefont{Niss}}, \bibinfo{journal}{J.
  Non-Crystal. Solids} \textbf{\bibinfo{volume}{407}}, \bibinfo{pages}{190}
  (\bibinfo{year}{2015}).

\bibitem[{\citenamefont{Ingebrigtsen
  et~al.}(2012{\natexlab{c}})\citenamefont{Ingebrigtsen, B{\o}hling,
  Schr{\o}der, and Dyre}}]{thermoscl}
\bibinfo{author}{\bibfnamefont{T.~S.} \bibnamefont{Ingebrigtsen}},
  \bibinfo{author}{\bibfnamefont{L.}~\bibnamefont{B{\o}hling}},
  \bibinfo{author}{\bibfnamefont{T.~B.} \bibnamefont{Schr{\o}der}},
  \bibnamefont{and} \bibinfo{author}{\bibfnamefont{J.~C.} \bibnamefont{Dyre}},
  \bibinfo{journal}{J. Chem. Phys.} \textbf{\bibinfo{volume}{136}},
  \bibinfo{pages}{061102} (\bibinfo{year}{2012}{\natexlab{c}}).

\bibitem[{\citenamefont{Ingebrigtsen et~al.}(2013)\citenamefont{Ingebrigtsen,
  Errington, Truskett, and Dyre}}]{ingebrigtsen2013}
\bibinfo{author}{\bibfnamefont{T.~S.} \bibnamefont{Ingebrigtsen}},
  \bibinfo{author}{\bibfnamefont{J.~R.} \bibnamefont{Errington}},
  \bibinfo{author}{\bibfnamefont{T.~M.} \bibnamefont{Truskett}},
  \bibnamefont{and} \bibinfo{author}{\bibfnamefont{J.~C.} \bibnamefont{Dyre}},
  \bibinfo{journal}{Phys. Rev. Lett.} \textbf{\bibinfo{volume}{111}},
  \bibinfo{pages}{235901} (\bibinfo{year}{2013}).

\bibitem[{\citenamefont{Ingebrigtsen and Dyre}(2014)}]{ingebrigtsen2014}
\bibinfo{author}{\bibfnamefont{T.~S.} \bibnamefont{Ingebrigtsen}}
  \bibnamefont{and} \bibinfo{author}{\bibfnamefont{J.~C.} \bibnamefont{Dyre}},
  \bibinfo{journal}{Soft Matter} \textbf{\bibinfo{volume}{10}},
  \bibinfo{pages}{4324} (\bibinfo{year}{2014}).

\bibitem[{\citenamefont{Separdar et~al.}(2013)\citenamefont{Separdar, Bailey,
  Schr{\o}der, Davatolhagh, and Dyre}}]{sllod}
\bibinfo{author}{\bibfnamefont{L.}~\bibnamefont{Separdar}},
  \bibinfo{author}{\bibfnamefont{N.~P.} \bibnamefont{Bailey}},
  \bibinfo{author}{\bibfnamefont{T.~B.} \bibnamefont{Schr{\o}der}},
  \bibinfo{author}{\bibfnamefont{S.}~\bibnamefont{Davatolhagh}},
  \bibnamefont{and} \bibinfo{author}{\bibfnamefont{J.~C.} \bibnamefont{Dyre}},
  \bibinfo{journal}{J. Chem. Phys.} \textbf{\bibinfo{volume}{138}},
  \bibinfo{pages}{154505} (\bibinfo{year}{2013}).

\bibitem[{\citenamefont{Veldhorst et~al.}(2014)\citenamefont{Veldhorst, Dyre,
  and Schr{\o}der}}]{veldhorst2014}
\bibinfo{author}{\bibfnamefont{A.~A.} \bibnamefont{Veldhorst}},
  \bibinfo{author}{\bibfnamefont{J.~C.} \bibnamefont{Dyre}}, \bibnamefont{and}
  \bibinfo{author}{\bibfnamefont{T.~B.} \bibnamefont{Schr{\o}der}},
  \bibinfo{journal}{J. Chem. Phys} \textbf{\bibinfo{volume}{141}},
  \bibinfo{pages}{054904} (\bibinfo{year}{2014}).

\bibitem[{\citenamefont{Schr{\o}der and Dyre}(2014)}]{thomas2014}
\bibinfo{author}{\bibfnamefont{T.~B.} \bibnamefont{Schr{\o}der}}
  \bibnamefont{and} \bibinfo{author}{\bibfnamefont{J.~C.} \bibnamefont{Dyre}},
  \bibinfo{journal}{J. Chem. Phys.} \textbf{\bibinfo{volume}{141}},
  \bibinfo{pages}{204502} (\bibinfo{year}{2014}).

\bibitem[{\citenamefont{B{\o}hling et~al.}(2012)\citenamefont{B{\o}hling,
  Ingebrigtsen, Grzybowski, Paluch, Dyre, and Schr{\o}der}}]{beyond}
\bibinfo{author}{\bibfnamefont{L.}~\bibnamefont{B{\o}hling}},
  \bibinfo{author}{\bibfnamefont{T.~S.} \bibnamefont{Ingebrigtsen}},
  \bibinfo{author}{\bibfnamefont{A.}~\bibnamefont{Grzybowski}},
  \bibinfo{author}{\bibfnamefont{M.}~\bibnamefont{Paluch}},
  \bibinfo{author}{\bibfnamefont{J.~C.} \bibnamefont{Dyre}}, \bibnamefont{and}
  \bibinfo{author}{\bibfnamefont{T.~B.} \bibnamefont{Schr{\o}der}},
  \bibinfo{journal}{New J. Phys.} \textbf{\bibinfo{volume}{14}},
  \bibinfo{pages}{113035} (\bibinfo{year}{2012}).

\bibitem[{\citenamefont{Pond et~al.}(2011)\citenamefont{Pond, Errington, and
  Truskett}}]{pond2011}
\bibinfo{author}{\bibfnamefont{M.~J.} \bibnamefont{Pond}},
  \bibinfo{author}{\bibfnamefont{J.~R.} \bibnamefont{Errington}},
  \bibnamefont{and} \bibinfo{author}{\bibfnamefont{T.~M.}
  \bibnamefont{Truskett}}, \bibinfo{journal}{J. Chem. Phys.}
  \textbf{\bibinfo{volume}{135}}, \bibinfo{pages}{124513}
  (\bibinfo{year}{2011}).

\bibitem[{\citenamefont{Veldhorst et~al.}(2015)\citenamefont{Veldhorst,
  Schr{\o}der, and Dyre}}]{veldhorst2015}
\bibinfo{author}{\bibfnamefont{A.~A.} \bibnamefont{Veldhorst}},
  \bibinfo{author}{\bibfnamefont{T.~B.} \bibnamefont{Schr{\o}der}},
  \bibnamefont{and} \bibinfo{author}{\bibfnamefont{J.~C.} \bibnamefont{Dyre}},
  \bibinfo{journal}{Phys. Plasmas} \textbf{\bibinfo{volume}{22}},
  \bibinfo{pages}{073705} (\bibinfo{year}{2015}).

\end{thebibliography}

\end{document}